\documentclass[%
 reprint,
superscriptaddress,
 amsmath,amssymb,
 aps,
pra,
]{revtex4-2}

\usepackage{graphicx}% Include figure files
\usepackage{dcolumn}% Align table columns on decimal point
\usepackage{bm}% bold math
\usepackage{hyperref}% add hypertext capabilities
\usepackage[mathlines]{lineno}% Enable numbering of text and display math
\usepackage{amsmath}
\usepackage{mathtools}
\usepackage{float}
\usepackage{amsfonts}
\usepackage{amssymb}
\usepackage{dcolumn} %% tables cols aligned at decimal point
\usepackage{array}
\newcolumntype{P}[1]{>{\centering\arraybackslash}p{#1}}
\newcolumntype{M}[1]{>{\centering\arraybackslash}m{#1}}
\newcolumntype{C}[1]{>{\centering\arraybackslwash}p{#1}}
\usepackage{float}
\usepackage{psfrag}
\usepackage{tabularx}
\usepackage{stackengine}
\usepackage{amssymb}
\usepackage{mathtools}  
\usepackage{xfrac} 
\usepackage[T1]{fontenc}
\usepackage{graphicx}
\usepackage{diagbox}
\usepackage[caption=false]{subfig}
\usepackage{tikz}
\usetikzlibrary{positioning}
\usetikzlibrary{arrows}
\usetikzlibrary{trees}
\usepackage{amssymb}
\usetikzlibrary{decorations.pathmorphing}
\usetikzlibrary{decorations.markings}
\usetikzlibrary{automata,positioning}
\usepackage{braket}
\usepackage{rotating}
\usepackage{adjustbox}
\usepackage{ragged2e}
\usepackage{simplewick}
\usepackage{simpler-wick}

\usepackage{csquotes}
\usepackage{physics,amsmath}
\usepackage{xcolor}
\usepackage{fancyhdr}
\UseRawInputEncoding
\usepackage[normalem]{ulem}

\usepackage{scalerel}
\usepackage{hyperref}
\usepackage{appendix}
\usepackage{multirow}
\usepackage[colorinlistoftodos]{todonotes}
\usepackage{chngcntr}
\usepackage{booktabs}
\usepackage{amsmath,amssymb}

\newcounter{algl}
\newcounter{algno}
\newcommand{\algln}[2][0]{\par\noindent\refstepcounter{algl}%
  \hangindent=\dimexpr 1.9em+#1em\relax\hangafter=1
  \makebox[1.5em][r]{\thealgl:}\hspace{0.4em}\hspace{#1em}#2}

\newcommand{\algkw}[1]{\textbf{#1}}
\newcommand{\algrule}{\par\vspace{2pt}\hrule height 0.4pt\vspace{3pt}}
\begin{document}

\preprint{APS/123-QED}

% \title{Physics-Informed Generative Machine Learning for Accelerated\\ Quantum-Centric Supercomputing}% Force line breaks with \\
% \title{Targeted Hilbert Space Exploration with Physics-Informed 
% Generative 
% Machine Learning
% % Driven Configuration Recovery
% toward Accelerated Quantum-Centric Supercomputing}% Force line breaks with \\
% \title{Quantum-Centric Supercomputing through Generative Machine Learning and
% Perturbation-Informed Targeted Hilbert Space Exploration}% Force
\title{
% Quantum-Centric Supercomputing for Extended Molecular Systems with Limited Shot Budget \\~\\
% Harnessing Physics-Informed Generative Machine Learning for Quantum-Centric Supercomputing of Extended Molecular Systems under Limited Shot Budgets \\~\\
% From Molecular Fragments to Extended Systems: Quantum-Centric Supercomputing under Limited Shot Budget through Physics-informed Generative Machine Learning\\~\\
% Distributed Quantum-Centric Supercomputing with Mapped Tensor Product States and Generative Learning for Extended Systems \\~\\
% Distributed Quantum-Centric Supercomputing with Generative Learning Driven Aggregation for Extended Systems \\~\\
Distributed Quantum-Centric Supercomputing with Multi-Layered Generative Learning for Intermolecular Interactions}
%line breaks with \\

\author{Chayan Patra}
\thanks{These authors contributed equally}
\affiliation{%
Department of Chemistry,  \\ Indian Institute of Technology Bombay, \\ Powai, Mumbai 400076, India
}%
\author{Prayas Ghosh}%
\thanks{These authors contributed equally}
\affiliation{%
Department of Chemistry,  \\ Indian Institute of Technology Bombay, \\ Powai, Mumbai 400076, India
}%
\author{Adwait Patkhedkar}
\affiliation{%
Department of Computer Engineering,  \\ Sardar Patel Institute of Technology, \\ Andheri, Mumbai 400058, India
}%
% \author{Sonaldeep Halder}%
% \affiliation{%
% Department of Chemistry,  \\ Indian Institute of Technology Bombay, \\ Powai, Mumbai 400076, India
% }%
\author{Rahul Maitra}
\email{rmaitra@chem.iitb.ac.in}
\affiliation{%
Department of Chemistry,  \\ Indian Institute of Technology Bombay, \\ Powai, Mumbai 400076, India
}%
\affiliation{Centre of Excellence in Quantum Information, Computing, Science \& Technology, \\ Indian Institute of Technology Bombay, \\ Powai, Mumbai 400076, India}
% % \collaboration{MUSO Collaboration}%\noaffiliation

%  % \homepage{http://www.Second.institution.edu/~Charlie.Author}
% \affiliation{
%  Second institution and/or address\\
%  This line break forced% with \\
% }%
% \affiliation{
%  Third institution, the second for Charlie Author
% \affiliation{%
%  Authors' institution and/or address\\
%  This line break forced with \textbackslash\textbackslash
% }%

% \collaboration{CLEO Collaboration}%\noaffiliation

\date{\today}% It is always \today, today,

\begin{abstract}

Quantum-centric supercomputing (QCSC) approaches, such as sample-based quantum diagonalization (SQD),
combine quantum processors with high-performance computers (HPC), offering a promising route toward solving challenging electronic
structure problems. However, implementation of QCSC approaches to large active space calculations is constrained by the associated number of qubits, deep quantum circuits,
sampling cost, and large-scale Hamiltonian diagonalization.
To overcome these limitations, we introduce a distributed QCSC framework that decomposes complex molecular problems into parallel sub-problems requiring fewer qubits, shallower circuits, and reduced shot budgets. A physically motivated, two-step aggregation scheme then reconstructs the global solution: first, a \textit{computational bit assignment} strategy optimizes a cost function to appropriately map local bit indices to global ones, yielding a \textit{mapped tensor product state} (MTPS); second, an \textit{inter-fragment correlation recovery module} restores missing cross-fragment interactions.
Herein, we employ a multi-layered generative machine learning architecture, starting with utilizing the recently proposed physics-informed generative SQD (PIGen-SQD) as a distributed QCSC solver,
followed by a secondary generative module to model configuration distributions and identify dominant sectors of the global Hilbert space.
% In this work we have used multi-layered generative machine learning: one with the recently proposed
% physics-informed generative machine-learning driven SQD (PIGen-SQD) as distributed QCSC solver followed by
% a generative learning driven inter-fragment recovery module to learn the configuration distribution and predict dominant sectors of the Hilbert space.
For water and formamide dimers, 
%($\mathrm{(H_2O)_2}$ and \textcolor{red}{\sout{$\mathrm{(CH_3NO)_2}$}} \textcolor{brown}{$(\mathrm{HCONH_2})_2$}), 
our method achieves up to $95\%$ reduction in the diagonalization dimension while enhancing energy accuracy by an order of magnitude over conventional SQD. Furthermore, application to the water hexamer ($\mathrm{(H_2O)_6}$) cage successfully decomposes a 120-qubit problem into six 20-qubit monomeric
subsystems, and stitches the monomers under a dimeric approximation, demonstrating excellent agreement with the
reference classical methods.

\end{abstract}

\maketitle

%\tableofcontents

\section{Introduction}

Accurate quantum-mechanical treatment of larger molecular systems with nontrivial electronic correlation effects is beyond the scope of classical mean-field theories and requires sophisticated high-level classical treatments with theories like coupled-cluster (CC)\cite{cc3, cc5, crawford2000introduction}, configuration-interaction (CI)\cite{sherrill1999configuration}, and the density matrix renormalization group (DMRG)\cite{white1992density}. However, the computational cost of such methods scales steeply with system size and thus they are not scalable. Quantum computing algorithms show immense promise to efficiently and accurately solve such electronic structure problems. Nevertheless, hardware noise in currently available quantum processing units (QPUs) affects computations, leaving quantum algorithms like quantum phase estimation (QPE) infeasible. For near-term applications, the community relies upon hybrid quantum-classical approaches such as the variational quantum eigensolver (VQE)~\cite{peruzzo2014variational, cerezo2021variational, bharti2022noisy}, projective quantum eigensolver (PQE)~\cite{stair2021simulating, patra2024projective, patra2024toward}, or the more recent quantum-centric supercomputing (QCSC)/ quantum selected configuration interaction (QSCI) frameworks~\cite{kanno2023quantum}. VQE- or PQE-like algorithms suffer from impractically long parameter optimization cycles and local traps~\cite{anschuetz2022quantum, cerezo2021variational} in the cost-function landscapes, and active research is ongoing in this field to mitigate the effects of such challenges~\cite{grimsley2019adaptive,smart2021quantum,Grimsley2023,ryabinkin2018qubit,bauman2019downfolding,matsuzawa2020jastrow,ryabinkin2020iterative,yordanov2020efficient,kowalski2021dimensionality,mondal2023development,halder2022dual,claudino2021improving,halder2023corrections,halder2023measurement,kowalski2023quantum,burton2023exact,burton2024accurate,halder2024noise,halder2023machine,patra2024projective,patra2024toward,patra2025energy, haidar2025non}. QCSC takes a different approach by performing state preparation and measurement to obtain electronic configuration samples from quantum hardware, followed by diagonalization on a classical processor to compute the desired observable.

In the context of quantum chemistry, QCSC is a class of selected configuration interaction (SCI) problem
that leverages a quantum computer as a sampling engine to identify the dominant sector of the Hilbert space that contain the solution
of the Schr{\"o}dinger equation. However, present day quantum processors suffer from noise due to the interaction with the environment and
often produce large number of error-prone samples that affect the computation of accurate energetics. The recently proposed
sample-based quantum diagonalization (SQD)\cite{robledo2025chemistry} addresses this issues by introducing a self-consistent configuration recovery
(S-CORE) protocol to denoise the error-prone samples based on the average orbital occupancy and it subsequently expands the subspace to include configurations from the unexplored sector of the Hilbert space followed by diagonalization.

The efficiency and accuracy of any generic configuration recovery depend on hardware-generated sample quality, which is governed by
hardware noise, quantum-state expressivity, and the available shot budget. For a system with \(M\) orbitals and \(P\) particles, the
configuration space scales as \(\binom{M}{P}\), implying combinatorial growth with system size. Even under ideal noiseless
conditions, statistically meaningful sampling according to the Born's interpretation
may require number of shots that roughly scales as \(\mathcal{O}\left(\binom{M}{P}\right)\),
limiting the scalability of sample-based quantum algorithms under restricted circuit-execution budgets.
% Hardware noise and the cost of preparing expressive quantum states further burden classical high-performance computing (HPC)
% resources in the QCSC workflow. This motivates fragmenting the system into chemically relevant subsystems that retain essential
% electronic correlations, enabling distributed QCSC calculations followed by classical post-processing to reconstruct the supersystem
% solution.
Furthermore, hardware noise, the challenges associated with preparing highly
expressive quantum states, along with restricted measurement budgets collectively increase the computational burden on classical high-performance computing (HPC) resources within the QCSC workflow.
These limitations motivate a fundamental question: can we avoid performing quantum state preparation and sampling for the entire system by partitioning it into chemically relevant subsystems that
retain nontrivial electronic correlation effects? Such an approach would enable the application of distributed QCSC solvers to individual sub-problems, followed by classical post-processing to aggregate the
sub-systems
for accurate solution of the super-system.
Specifically, partitioning an \(M\)-qubit system into \(N\) fragments, with \(M_k\) qubits and \(P_k\) electrons in
\(k\)-th fragment, reduces the sampling cost to approximately \(\mathcal{O}\left(\binom{M_k}{P_k}\right)\) per fragment. Fragmentation also reduces noisy entangling-gate counts and circuit depth, thereby enhancing the quality of
samples per fragment.
Toward this, there are many recent quantum-classical hybrid algorithmic developments that focus on combining QCSC, particularly
SQD with existing classical embedding strategies to treat extended systems
such as SQD with density matrix
embedding theory (DMET-SQD)\cite{shajan2025toward, patra2026quantum}, SQD with localized active space
approximation (LASSQD)\cite{wang2026localized}, and several other works\cite{shajan2026molecular,merz2026crossing, yamamoto2026quantum}. The above-mentioned methods uses SQD as an active-space sub-system solver and
are recombined by embedding them within each other's environment at a lower level of theory.
% A widely adopted strategy to alleviate this bottleneck exploits the locality of electronic interactions through
% fragmentation and embedding techniques, wherein a large molecular system is partitioned into a collection of smaller, overlapping or non-overlapping subsystems that are treated using QCSC solvers
% while the surrounding environment is described at a computationally less demanding level.
% \cite{knizia2012density,wouters2016practical}

In this work, we take a complementary approach to enable larger active space QCSC simulations even with
limited shot budget and reduced diagonalization cost.
Here we propose a distributed QCSC approach that involves decomposition of an extended
molecular system into chemically motivated fragments that are solved using parallel fragment QCSC solvers
followed by a physically motivated aggregation step.
To this end, the recently developed physics-informed generative machine-learning driven SQD (PIGen-SQD)\cite{Patra_2026}
is explored as the parallel fragment QCSC solver which proposes a concerted use of quantum hardware samples, low order perturbation
theory and generative machine learning (Gen-ML) to explore the dominant sectors of the Hilbert space.
PIGen-SQD has been demonstrated to provide extremely accurate
description of the electronic strong correlation effects with the associated diagonalization performed
in drastically compact subspaces compared to SQD. While distributed PIGen-SQD solvers provide
accurate description of the intra-fragment correlations, the solution of the super-system requires accurate
integration of these fragment solutions along with the recovery of inter-fragment correlations.
This is done with a two-fold aggregation stage: first, a product state construction from the dominant fragment
configurations is performed, which is assumed to provide non-trivial support to the eigenstate of the super-system Hamiltonian.
However, the product state requires appropriate one-to-one assignment of the local computational
bits to that of the super-system such that dominant local configurations are properly mapped to the
canonical fock space. Toward this, we have proposed a \textit{computational bit mapping} strategy here
which solves a \textit{linear assignment problem} (LAP) for the local-to-global computational bit correspondence.
With this mapping strategy, one can construct a \textit{mapped tensor product state} (MTPS) directly from the
local QCSC solutions. MTPS provides a good initialization to the second stage of aggregation which involves
an \textit{inter-fragment correlation recovery} stage. In this paper we have explored an additional layer of Gen-ML to entangle two or more subsystems and reconstruct the bitstrings in the n-meric 
computational basis. The algorithm is referred to as
% \textcolor{blue}{distributed PIGen-SQD or Dist-PIGen-SQD},
PIGen-SQD \textit{with mapped aggregation
of partitioned subspace} or PIGen-SQD(MAPS). Additionally, to provide a fair comparison, we have considered
the traditional S-CORE mechanism initialized with MTPS for the inter-fragment recovery as well.
PIGen-SQD(MAPS) provides a few key advantages over traditional SQD performed over
the entire system: (1) it can be performed with limited shot budget since quantum circuits are only run
over fragments, (2) the diagonalization dimension is significantly low while energy accuracy 
with the reference is substantially better in comparison, and (3) depending on the complexity of
the problem and the computational resources available, one can choose the number of monomeric
fragments ($n$) that needs to be entangled explicitly during the inter-fragment correlation 
recovery.

Here we have used PIGen-SQD(MAPS) to study non-covalent inter-molecular interactions 
which plays critical roles in biological processes such as protein-folding, membrane assembly, 
and is absolutely essential for drug discovery pipelines. Particularly, we have studied
water dimer, which is the prototypical system for studying hydrogen bonds, and 
formamide (FMA) dimer which is biochemically relevant as the smallest system containing the peptide bond.
Our numerical studies on the water and formamide dimers demonstrate that PIGen-SQD(MAPS) achieves up to 95\% reduction in the diagonalization space and an order-of-magnitude improvement in energy accuracy, while requiring only a fraction of the shots compared to conventional SQD performed on the dimeric system.
% We have further tested PIGen-SQD(MAPS) on $\mathrm{(H_2O)}_6$ with a cage structure with fragment-pair $(n=2)$ approximation that distributes 120 qubit active
% space problem requiring hardware samples from 20 qubit PQC, followed by an inter-fragment 
% correlation recovery over dimeric supersystem followed by a many-body expansion step for estimating the energy of the hexamer.
We further benchmark PIGen-SQD(MAPS) on cage-shaped $\mathrm{(H_2O)}_6$ using a fragment-pair ($n$=2) approximation, partitioning the 120-qubit active-space problem into subsystems requiring
hardware samples from 20-qubit quantum circuits. Inter-fragment correlations are subsequently
recovered at the dimeric supersystem level, followed by a many-body expansion to estimate the total energy of the
hexamer. Importantly, in our algorithm, the quantum 
hardware is employed to sample only for the monomeric units that significantly improves the 
probability of obtaining high quality samples with limited shot budget.

\section {Treatment of Extended Systems within QCSC Framework: General Considerations}

QCSC algorithms such as SQD, proposes a quantum hardware as an efficient engine to sample from dominant sector of the Hilbert space
for accurate solution of the fermionic Hamiltonian
\begin{equation} \label{eq: fermionic hamiltonian second quantized}
    \hat{H} = \sum_{pq} h_{pq} \hat{a}^{\dagger}_p \hat{a}_q + \sum_{pqrs} v_{pqrs} \hat{a}^{\dagger}_p \hat{a}^{\dagger}_q \hat{a}_s \hat{a}_r
\end{equation}
where, $p,q,r,s$ are molecular orbital indices, $h_{pq}$ and $v_{pqrs}$ are respectively the one and two electron integrals in the molecular orbital basis
and $\hat{a}^{\dagger}_p,\hat{a}_q$ are the creation and annihilation operators.
% and S-CORE mechanism to accurately identify the dominant sector of the Hilbert 
% space that has substantial support to the target state. The Hamiltonian is then projected onto this sector 
% of the Hilbert space and is diagonalized to solve the associated selected CI problem.
% QCSC/ QSCI, is a class of hybrid quantum-classical algorithms that outsources this classically intractable core-space selection to a quantum processor, followed by classical projection and diagonalization of the Hamiltonian in the selected space for energy estimation.
% This requires a fairly accurate fermionic state preparation on a quantum hardware.
This is achieved by mapping the fermionic electronic-structure problem into the quantum-computing framework by encoding 
the $M$ molecular spinorbitals to $M$ qubits, so that each electronic configuration
corresponds to a computational basis state (or configuration) $\ket{\phi} = \ket{x_M..x_2 x_1}$ with $(x_M...x_2x_1) \in \{0,1\}^M$, where the single-qubit 
states $\ket{0}$ and $\ket{1}$ represent unoccupied and occupied spinorbitals respectively. The
Jordan-Wigner\cite{jordan1993paulische} transformation maps the second-quantized fermionic operators to qubit operators. A trial 
state on $M$ qubits is prepared by
a physically motivated ansatz which is expressed by a parameterized quantum circuit (PQC).
Ideally the state preparation should be done in a way that it has a substantial overlap with the target state.
In this work, we have used the local unitary cluster jastrow (LUCJ)\cite{motta2023bridging}
ansatz in its single-layer form,
\begin{equation} \label{LUCJ}
    \ket{\Psi}_{\mathrm{LUCJ}} = e^{-\hat{K}_2} e^{\hat{K}_1} e^{i\hat{J}_1} e^{-\hat{K}_1} \ket{\Phi_0}
\end{equation}
where $\hat{K}_1$ and $\hat{K}_2$ are one-body operators and $\hat{J}_1=\sum_{p\sigma r \tau}J_{p\sigma,r\tau}\hat{n}_{p\sigma}\hat{n}_{r\tau}$ is a density-density operator with $\hat{n}$ being
the number operator, and $\ket{\Phi_0}$ is the Hartree-Fock (HF) configuration. Repeated measurement of this circuit in the computational basis produces the distribution of outcomes
\begin{equation}
    \tilde{\chi} = \{ \phi | \phi \sim \tilde{P}(\phi) \}
\end{equation}
with the associated noisy probabilities
\begin{equation}
    \tilde{P}(\phi) = \frac{N_{{\phi}}}{N_{\mathrm{shots}}}
\end{equation}
where $N_{{\phi}}$ is the frequency of configuration ${\ket{\phi}}$ and $N_{\mathrm{shots}}$ is the total number of shots.
Due to the noise present in present-day quantum processors, the sampled distribution is contaminated with particle number and $z$ component of spin ($\hat{S}_z$) non-conserving configurations
and requires subsequent classical post-processing to correct the noise-affected distribution. 
The original SQD formulation involves the S-CORE post-processing module that is based on average occupancy 
of the associated spinorbitals and selected CI loops to identify dominant configurations for the Hamiltonian diagonalization.
The performance of the S-CORE module, in terms of both computational efficiency and accuracy, depends heavily on the quality of the hardware-generated samples, which is influenced by the
prevailing hardware noise, the expressivity of the prepared quantum state, and the available shot budget.
These parameters pose significant challenges to extend QCSC algorithms to extended systems of practical interest. 

From a quantitative point of view, the Schr{\"o}dinger's equation of an extended system (with $M$ orbitals and $P$ electrons)
has the general CI solution
\begin{equation}
    \ket{\Psi} = \sum_g c_g \ket{\Phi_g}
\end{equation}
where, $\ket{\Phi_g} \in \{0,1\}^{M}$ are computational bases and $c_g$ are the associated coefficients in the
linear expansion. Solving the corresponding QCSC problem involves PQC with $M$ qubits (assuming one-to-one
spinorbital-to-qubit mapping), $\mathcal{O}(M^2)$ two qubit gate complexity per layer (considering UCJ circuit)\cite{belagali2026efficientclassicalsimulationlargescale},
and at least $\mathcal{O}(\binom{M}{P})$ shot budget for relevant sampling statistics,
which scales unfavorably with system size.
Alternatively, a more tractable approach would be to partition the entire Hamiltonian into
\begin{equation}
    \hat{H} = \hat{H}^{(0)} + \hat{H}^{\prime}
\end{equation}
where, $\hat{H}^{(0)}$ has an additively separable form as it contains only intra-fragment terms and mean-field inter-fragment effects, and is exactly solvable. The residual non-local inter-fragment interaction terms
are considered into $\hat{H}^{\prime}$.
In general, if the extended system can be partitioned into $N$ fragments, the zeroth order Hamiltonian has the generic form
\begin{equation}
    \hat{H}^{(0)} = \sum_{k=1}^{N} \hat{H}_k^{(0)}
\end{equation}
% \textcolor{red}{I think it is better to carry the sub/superscript 0 to identify it as zeroth order Hamiltonian and getting inconsistent with the equation 5 and 7.}
where $k$ is the fragment index. The local fragment Hamiltonian is defined as
% $\hat{H}_k = \sum_{i \in I} \hat{h}_k(i) + \sum_{ij \in I} \hat{V}_k ({ij}) + \hat{V}_k^{(eff)} $
\begin{equation}
\hat{H}_k^{(0)} = \sum_{I \in k} \hat{h}_k^{(0)}(I) + \sum_{IJ \in k} \hat{V}_k^{(0)} ({IJ}) + \hat{V}_k^{(0){\mathrm{(eff)}}}     
\end{equation}
consisting of one-body ($\hat{h}^{(0)}_k(I)$), two-body ($\hat{V}^{(0)}_k ({IJ})$) intra-fragment operators and an effective potential $\hat{V}_k^{(0){(\mathrm{eff})}}$
that encodes the mean-field effects of all other fragments on $k$-th fragment.
% Here $I$-th fragment is the smallest unit
% having $K_I$ orbitals and $N_I$ particles.
The dimension of local Hilbert space $\mathcal{H}_k$ for $k$-th fragment having $M_k$ spinorbitals and $P_{k}$ number of
particles is $\binom{M_k}{P_k}$. Assuming the fragmentation is done in such a way that the local Hilbert spaces
are computationally challenging and involves non-trivial electronic strong correlation effects, one may solve each fragment with efficient QCSC solvers that provide highly accurate
solutions for the fragments.
The ground state of the fragment CI problems $\hat{H}^{(0)}_k \ket{\Psi^{(0)}_k} = E_k^{(0)} \ket{\Psi^{(0)}_k}$
obtained from QCSC solvers are represented as a linear combination
% can be approximated by a linear combination of local fragment configurations
\begin{equation}
    \ket{\Psi^{(0)}_k} = \sum_{\alpha} c_{\alpha_k}^{(0)} \ket{\phi_{\alpha_k}^{(0)}}
\end{equation}
where, $\ket{\phi_{\alpha_k}^{(0)}} \in \{0,1\}^{M_k}$ are the computational bases (configurations) of the $k$-th fragment.
This eigenvalue equation for each fragment can be solved by QCSC solvers in parallel QPU-HPC platforms to provide the associated eigenstate solutions which consist of a set of configurations and the associated CI coefficients.
Under an $n$-meric approximation with $n \leq N$, 
% This leads to
the product-state solution of the zeroth order $n$-meric Hamiltonian is
% $\hat{H}^{(0)}$
\begin{equation} \label{eq: direct product state}
    \ket{\Psi^{(0)}} = \prod_{k=1}^{n} \ket{\Psi^{(0)}_k} 
\end{equation}
Note that this product-state solution is obtained by enforcing spin ($z$ component of the spin angular momentum $\hat{S}_z$) and particle number conservation over each fragment
as this constraint is necessary (at least in the zeroth order solution level) for efficient QCSC implementation.
% Similarly the general solution of the composite system (eigenstate of the full Hamiltonian $\hat{H}$) can be expressed as
% \begin{equation}
%     \ket{\Psi} = \sum_g c_g \ket{\Phi_g}
% \end{equation}
% where, $\ket{\Phi_g} \in \{0,1\}^{M}$ are computational basis of the $n$-meric composite system with $M$ being the associated number of spinorbitals.

At this stage one can start with the product-state solution $\ket{\Psi^{(0)}}$ (Eq. \eqref{eq: direct product state}) which neglects
a substantial part of the inter-fragment correlation effects, and then treat the effects of $\hat{H}^{\prime}$ perturbatively
to get the solution of the full extended system. The recently proposed FragPT2 method \cite{koridon2025fragpt2} takes such an approach in a classical computing platform, where
the zeroth order solution is supplemented by a perturbative treatment of the inter-fragment correlation effects.
% \textcolor{blue}{ADD SOME MORE REF HERE}
While such a perturbative treatment is efficient and provides quantitative description of charge transfer, dispersion and triplet-triplet coupling effects, its accuracy might get limited due to the perturbative treatment 
of the inter-fragment effects. In order 
to obtain chemically meaningful accuracy, one may take a non-perturbative approach like (Full) CI over a 
suitably constructed n-meric basis. This way, the quality of the solution primarily depends on the fragmentation scheme and not on the insufficient treatment of correlation.
To proceed along this line within a QCSC/QSCI framework, one
needs to project the n-meric Hamiltonian onto a compact subspace of the corresponding $n$-meric Hilbert space
and solve the associated CI diagonalization problem. In such case, the computational bottleneck primarily comes from the 
large $n$-meric diagonalization and one can account for only low order n-meric approximation among the fragments. Such an approach presumes that the solutions of the local monomeric CI problems have 
substantial support to the solution of the composite $n$-meric system.
Consequently, one can construct a dominant 
subspace in the canonical Hilbert space directly from the information of the set of (dominant) local computational 
bases over various fragments.
This assumption requires appropriate mapping of local computational basis originating from 
fragment QCSC solvers to the global spin-particle conserved fock space of the canonical system.
This directly transfers the local fragment characteristics to the global Hilbert space.  
However, such mapping is a non-trivial task, particularly if there are near degenerate orbitals originating 
from different fragments, as it requires appropriate one-to-one assignment of 
local (fragment) molecular orbitals to the canonical molecular orbitals
of the entire system. In this context, it is important to clarify that,
the terms \textit{supersystem}, \textit{canonical system}, and \textit{global subspace} are used interchangeably throughout this work to denote the $n$-meric system,
and the associated Hilbert space.
% the terms supersystem, canonical system and global subspace will be interchangeably used
% to represent $n$-meric system.

Here we address this by presenting a \textit{computational bit assignment} strategy that leads to appropriate
reconstruction of the global basis from local configurations.
Since for chemical systems computational bits are directly associated with molecular spinorbitals,
in this case the mapping is done by analyzing orbital overlap and spectral characters of super-system and sub-system orbitals.
% orbital mapping and computational bit
% mapping are synonymous
% \textcolor{red}{(While this is absolutely true, I think this slightly trivializes it. Better that we say "Such computational bit mapping is performed in the orbital basis and then mapped to the canonical bit strings" or something like this...)}
% and will be used interchangeably.
The assignment information are subsequently used to map the global basis from the dominant local basis
and such a reconstruction is referred to as \textit{mapped tensor product state} (MTPS).
% The computational bit mapping thus leads to
% a product state construction, which we refer to as \textit{mapped tensor product state} (MTPS).
Computationally, MTPS forms a set of
bitstrings that provides us with the optimal basis with appropriate distribution of occupancy and spin properties
to initiate \textit{inter-fragment correlation recovery} for the $n$-meric
super-system.
One may note that the mapped product state, as discussed 
above, is structurally different from the direct product state $\ket{\Psi^{(0)}}$ in Eq. 
\eqref{eq: direct product state}. With an accurately designed bit mapping, this modified product space forms a substantially compact 
subspace in the composite Hilbert space which is purely constructed based on the knowledge of the local 
CI solutions. Such an approach would drastically reduce the complexity of the QCSC problem if
the fragmentation is done in a way such that $\binom{M_k}{P_k} \ll \binom{M}{P}$ for all $k$.
This workflow enables one to solve a set of 
tractable monomeric QCSC problems and reconstruct a dominant subspace of the canonical Hilbert space
that has non-trivial support to the target eigenroot.

In this aspect, it is important to mention that, there is no unique way to construct the monomeric Hamiltonian. Furthermore, the choice of the many-body
expansion order $n$ 
may typically be user defined depending on the complexity of the system and the computational budget at disposal. 
In Section \ref{sec: Methodological Development}, we will discuss the theoretical and algorithmic development
that defines molecular fragmentation and effective Hamiltonian construction, parallel QCSC solvers, 
computational bit assignment strategy and MTPS construction, and inter-fragment
correlation recovery protocols toward accurate treatment of extended chemical systems.
% framework used for
% computational bit mapping strategy and MTPS construction.

% Before moving into the details of computational bit mapping strategy and MTPS construction, we will first discuss briefly the PIGen-SQD framework which will be used
% as parallel fragment solvers to obtain accurate local CI solutions.

% CBE formalism that provides the foundation of the same.

% \section{Methodological Development} \label{subsec: extended system treatment}
\section{Problem Decomposition and Distributed QCSC Workflow} \label{sec: Methodological Development}

\subsection{Hamiltonian Partitioning and Molecular Fragmentation} \label{subsec: hamiltonian partitioning and fragmentation FMO}

Molecular fragmentation provides a systematic route to reducing the computational cost of electronic-structure
calculations by decomposing an extended molecular system into smaller, chemically meaningful subsystems. Among the established
approaches, the fragment molecular orbital (FMO) method
partitions a molecular system into $N$ fragments and evaluates fragment, fragment-pair, ... up to n-meric fragment
contributions to approximate the properties of the entire system \cite{Kitaura1999}. Similarly, energy-based fragmentation methods
express the total energy in terms of contributions from subsystems, often using a many-body expansion or related energy
decomposition schemes \cite{Collins2015}. These approaches offer a natural framework for quantum-centric supercomputing (QCSC),
wherein individual fragments can be assigned to quantum-classical solvers and processed independently or in parallel, followed by
classical reconstruction of the properties of the full system. For systems in which strong electronic correlation and fragment
environment entanglement are important, density matrix embedding theory (DMET) provides a complementary strategy by embedding a
fragment in an effective bath designed to capture its coupling to the surrounding system \cite{Knizia2012,Knizia2013}. Adapting
these fragmentation and embedding principles to QCSC could enable resource-intensive quantum calculations to focus on the most
chemically relevant subsystems, while treating interfragment interactions through an appropriate classical or embedding-based
correction scheme.
% For practically relevant extended systems, it is impractical to implement QSCI over the entire molecule even with the MTPS-QSCI
% algorithm discussed so far as the dimension of the corresponding dominant subspace \textcolor{red}{(YOU MEAN OF THE CANONICAL SYSTEM?)} increases excessively. Various molecular fragmentation approaches
% such as Fragment Molecular Orbital (FMO) theory in this aspect are extremely useful which divides the large molecular systems into chemically important
% sub-systems and treats each fragment separately in a higher level of theory.

In this paper, we have used FMO as the fragmentation tool, however, one may choose other relevant fragmentation
mechanisms as well.
FMO with $n$-body approximation (known as FMO-n) involves the following steps: (i) partitioning the entire system into
N fragments (monomers),
(ii) monomeric calculation, dimeric  calculation $\cdots$ up to $n$-meric calculation (where $n \leq N$) with a given method, and (iii) total property evaluation.
For example, under the FMO-2-RHF approximation, a large molecular system is divided into individual fragments.
The molecular orbitals (MOs) of each fragment are self consistently optimized in the field of external potential created
by the surrounding ($\mathrm{(N-1)}$)
fragments. The Hamiltonian for the monomer is given by,
%\begin{equation}
%    \hat{H}_I \Psi_I = E_I \Psi_I
%\end{equation}
\begin{equation}
\begin{split}
        & \hat{H}_k = \sum_{I\in k} \Big[ -\frac{1}{2} \nabla_I^2 - \sum_A \frac{Z_A}{|r_I - R_A |} + \sum_{l \neq k} \int dr' \frac{\rho_l(r')}{|r_I - r'|} \Big] \\
        & + \sum_{I \in k} \sum_{I < J \in k} \frac{1}{|r_I - r_J|}
\end{split}
\end{equation}
to solve the monomeric problems: $\hat{H}_k \Psi_k = E_k \Psi_k$. Here,
$k, l$ and $r_I, R_A$ represent different monomeric fragments and electronic and nuclear coordinates respectively, $\nabla_I^2$ is the kinetic energy operator, $Z_A$ is the atomic number and $\rho_l(r')$ denotes the electronic density at position $r'$ in $l$-th fragment.
Similarly the MOs of a fragment pair (dimer) are obtained self-consistently in the field produced by the leftover
$\mathrm{(N-2)}$ fragments. The associated Hamiltonian is given by,
%\begin{equation}
%    \hat{H}_{IJ} \Psi_{IJ} = E_{IJ} \Psi_{IJ}
%\end{equation}
\begin{equation}
\begin{split}
        & \hat{H}_{kl} = \sum_{I\in k, l} \Big[ -\frac{1}{2} \nabla_I^2 - \sum_A \frac{Z_A}{|r_I - R_A |} + \sum_{m \neq k, l} \int dr' \frac{\rho_m(r')}{|r_I - r'|} \Big] \\
        & + \sum_{I \in k, l} \sum_{I < J \in k, l} \frac{1}{|r_I - r_J|}
\end{split}
\end{equation}
which is used to solve each dimeric problem: $\hat{H}_{kl} \Psi_{kl} = E_{kl} \Psi_{kl}$.
With the monomer and dimer calculations, one can combine the solutions to obtain the properties of the full system through the many-body expansion (MBE),
\begin{equation} \label{eq: many body expansion RHF}
    % E^{\mathrm{FMO2-RHF}} = E^{\mathrm{FMO1-RHF}} + \Delta E^{\mathrm{FMO2-RHF}} %E^{\mathrm{FMO1-RHF}} + \Delta E^{\mathrm{FMO2-RHF}}
    E^{\mathrm{FMO2-RHF}} = \sum_k E_k + \sum_{k<l} \Big[  E_{kl} - E_k - E_l \Big],
\end{equation}
% \begin{equation}
%     E^{\mathrm{FMO1-RHF}} = \sum_k E_k,
% \end{equation}
% \begin{equation}
%     \Delta E^{\mathrm{FMO2-RHF}} = \sum_{k<l} \Big[  E_{kl} - E_k - E_l \Big]
% \end{equation}
where, $E_k$ and $E_{kl}$ are electrostatically embedded HF energies of monomer and dimers respectively.
Since FMO-RHF lacks the description of explicit electronic correlation, QSCI methods can be implemented along with FMO approximation for accurate
correlation energy calculation with the corresponding MBE of the form
\begin{equation} \label{eq: many body expansion QSCI}
    E^{\mathrm{FMO2-QSCI}} = \sum_k E_k^{\mathrm{QSCI}} + \sum_{k<l} \Big[ E_{kl}^{\mathrm{QSCI}} - E_k^{\mathrm{QSCI}} - E_l^{\mathrm{QSCI}} \Big]
    % E^{\mathrm{FMO2-QSCI}} = E^{\mathrm{FMO2-RHF}} + E^{\mathrm{FMO2-corr}}, %\sum_k E_k^{\mathrm{QSCI}} + \sum_{k<l} \Big[ E_{kl}^{\mathrm{QSCI}} - E_k^{\mathrm{QSCI}} - E_l^{\mathrm{QSCI}} \Big]. %E^{\mathrm{FMO1-RHF}} + \Delta E^{\mathrm{FMO2-RHF}} 
\end{equation}
% \begin{equation}
%     E^{\mathrm{FMO2-corr}} = E^{\mathrm{FMO1-corr}} + \Delta E ^{\mathrm{FMO2-corr}},
% \end{equation}
% \begin{equation}
%     E^{\mathrm{FMO1-corr}} = \sum_k E_k^{\mathrm{corr}}
% \end{equation}
% \begin{equation}
%     \Delta E ^{\mathrm{FMO2-corr}} =  \sum_{k<l} \Big[ E_{kl}^{\mathrm{corr}} - E_k^{\mathrm{corr}} - E_l^{\mathrm{corr}} \Big]
% \end{equation}
So, within the FMO-2 approximation, monomer energies can be obtained with distributed QCSC solvers in parallel. Subsequently, to get the dimeric energy contributions ($E_{kl}^{\mathrm{QSCI}}$),
MTPS is constructed, followed by systematic retrieval of inter-fragment electronic correlation
% with configuration recovery modules.
with the help of
% inter-fragment correlation recovery with
Gen-ML or S-CORE driven modules
discussed in Section \ref{sec:inter-frag corr recovery}. Note that, in this case, no quantum hardware samples for dimeric system is required as it is entirely initialized by MTPS.
These steps can be generalized for arbitrary n-meric FMO approximations as well, albeit at an increased computational cost.

% \section{Physics-Informed Generative Sample-Based Quantum Diagonalization}
% \label{sec:pigen_sqd}

\subsection{PIGen-SQD Framework as Distributed Fragment QCSC Solver} \label{subsec: pigen-sqd}

PIGen-SQD is a class of efficient QCSC algorithms that uses quantum hardware samples, perturbative 
measures and generative machine learning (Gen-ML) to accurately identify the dominant sector of the Hilbert 
space that has substantial support to the target state.
The initial configuration space is constructed by combining configurations obtained from quantum hardware measurements with additional determinants selected using perturbative estimates.
% For hardware sampling, a parameterized quantum circuit $\hat{U}(\boldsymbol{\theta})$ acts on a reference state, typically the Hartree-Fock determinant, $\ket{\Psi(\boldsymbol{\theta})}=
% \hat{U}(\boldsymbol{\theta})\ket{\Phi_0}.$
% % \begin{equation}
% %     \ket{\Psi(\boldsymbol{\theta})}
% %     =
% %     \hat{U}(\boldsymbol{\theta})\ket{\Phi_0}.
% %     \label{eq:pigen_trial_state}
% % \end{equation}
% Repeated measurements in the computational basis generate a set of bitstrings corresponding to electronic configurations.
Since hardware noise can produce configurations that violate the required particle-number and spin-projection symmetries, the measured configurations are filtered to retain only those belonging to the desired symmetry sector.

To complement the hardware samples, many-body perturbation theory (MBPT) is employed to identify additional configurations with potentially significant contributions to the target state. Starting from the HF reference, the second-order M\o ller-Plesset (MP2) amplitudes are given by $    t_{ij}^{ab}
    =
    \frac{\langle ab\Vert ij\rangle}
    {\epsilon_i+\epsilon_j-\epsilon_a-\epsilon_b},$
% \begin{equation}
%     t_{ij}^{ab}
%     =
%     \frac{\langle ab\Vert ij\rangle}
%     {\epsilon_i+\epsilon_j-\epsilon_a-\epsilon_b},
%     \label{eq:pigen_mp2_amplitude}
% \end{equation}
where $i,j$ and $a,b$ denote occupied and virtual spin-orbitals, respectively, and $\epsilon_p$ is the energy of spin-orbital $p$. These amplitudes provide perturbative estimates of the importance of doubly excited determinants. Higher-rank configurations can subsequently be identified through hierarchical perturbative constructions involving connected contractions of quasi two-body excitation operators (called scattering operators) and MP2 amplitudes.
Explicit evaluation of all the tensor contractions required for higher-rank excitations can be computationally expensive. PIGen-SQD therefore employs sparse integral screening and index-based symbolic operations to identify candidate configurations without explicitly evaluating all possible tensor contractions. The resulting perturbatively selected configurations are combined with the symmetry-filtered hardware samples to form an enriched initial configuration space,
\begin{equation}
    \mathcal{B}^{(0)}
    =
    \mathcal{B}_{\mathrm{hw}}
    \cup
    \mathcal{B}_{\mathrm{MBPT}},
    \label{eq:pigen_initial_space}
\end{equation}
where $\mathcal{B}_{\mathrm{hw}}$ and $\mathcal{B}_{\mathrm{MBPT}}$ denote the hardware-derived and perturbatively selected determinant sets, respectively. The perturbative excitation rank and screening thresholds control the size of the initial space and the computational effort required for its construction.
% \subsubsection{Self-consistent configuration recovery}
% \label{subsec:pigen_recovery}
The molecular Hamiltonian of interest is projected onto the current determinant space $\mathcal{B}^{(m)}$ and diagonalized classically,
\begin{equation}
    \mathbf{H}^{(m)}\mathbf{c}^{(m)}
    =
    E^{(m)}\mathbf{c}^{(m)},
    \label{eq:pigen_projected_eigenproblem}
\end{equation}
where $\mathbf{H}^{(m)}$ is the Hamiltonian matrix in the current configuration space, $E^{(m)}$ is the target eigenvalue, and $\mathbf{c}^{(m)}$ is the corresponding CI coefficient vector. The coefficients are used to distinguish configurations with appreciable contributions to the target state from those with negligible contributions.
For a prescribed coefficient threshold $\epsilon_{\mathrm{coeff}}$, the retained dominant determinant set is defined as
\begin{equation}
    \mathcal{B}_{\mathrm{dom}}^{(m)}
    =
    \left\{
    \ket{\Phi_\mu}\in\mathcal{B}^{(m)}
    :
    |c_\mu^{(m)}|>\epsilon_{\mathrm{coeff}}
    \right\}.
    \label{eq:pigen_dominant_set}
\end{equation}
where, $\mu$ is the composite hole-particle index.
Configurations with coefficients below the threshold are recorded in a blacklist,
$\mathcal{B}_{\mathrm{blk}}^{(m)}$, to prevent their repeated selection during subsequent iterations. The dominant configurations are then used to construct the training distribution for the generative model.
Restricted Boltzmann Machine (RBM) (see Appendix \ref{appendix: RBM}) is used as the generative model in this case to learn the distribution of configurations from the CI-weighted training data
and to predict further dominant determinants from the unseen sectors of the Hilbert space.
For a bias-free training,
HF determinant is excluded from the RBM training data since the associated coefficient is usually several orders-of-magnitude higher than other configurations.
% to avoid biasing the learned distribution toward the reference configuration.
RBM is trained on this distribution and generates new candidate configurations through Gibbs sampling. The newly generated configurations are screened against the blacklist and combined with the existing determinant space to produce an expanded configuration space,
\begin{equation}
    \mathcal{B}^{(m+1)}
    =
    \mathcal{B}_\mathrm{dom}^{(m)}
    \cup
    \left(
    \mathcal{B}_{\mathrm{new}}^{(m)}\setminus\mathcal{B}_\mathrm{blk}^{(m)}
    \right),
    \label{eq:pigen_space_update}
\end{equation}
where $\mathcal{B}_{\mathrm{new}}^{(m)}$ denotes the set of newly generated configurations after enforcing the required physical symmetries and removing blacklisted determinants. The Hamiltonian is subsequently re-diagonalized in the expanded space, and the resulting CI coefficients provide updated training data for the next iteration.

The procedure alternates between RBM training and configuration generation, referred to as the \textit{micro-cycle}, and Hamiltonian diagonalization and configuration-space updating, referred to as the \textit{macro-cycle}. The iterative recovery is terminated when no new configurations are generated or when the energy change between successive macro-cycles satisfies
\begin{equation}
    \left|E^{(m)}-E^{(m-1)}\right|
    < \epsilon_E,
    \label{eq:pigen_energy_convergence}
\end{equation}
where $\epsilon_E$ is the prescribed energy convergence threshold. Algorithmic procedure for PIGen-SQD is summarized in Algorithm \ref{alg:pigen-sqd}.
The converged eigenvalue and eigenvector are obtained by diagonalizing the Hamiltonian in the final recovered configuration space.
The principal advantage of PIGen-SQD is the complementary role of its three components: quantum hardware provides samples from the prepared electronic state that are stochastically drawn irrespective of the excitation rank, perturbative screening introduces physically motivated configurations that may be absent from the measured distribution and provides an anchoring effect for the subsequent Gen-ML training, and generative learning facilitates iterative exploration of the dominant region of the symmetry-preserving Hilbert space. This combination aims to improve the accuracy of the recovered electronic state while controlling the dimension of the classical diagonalization problem.

% ---------- figure block ----------
\begin{figure}[t]
\setcounter{algl}{0}\small\raggedright\setlength{\parindent}{0pt}
\hrule height 0.6pt\vspace{3pt}
\refstepcounter{algno}\noindent\textbf{Algorithm \thealgno} PIGen-SQD\label{alg:pigen-sqd}
\algrule
\algln{\algkw{Input:} Hamiltonian $\hat H$, circuit $\hat U(\boldsymbol\theta)$, shots $N_{\rm shots}$, perturbative rank $n$, thresholds $\epsilon_{\rm int}$, $\epsilon_{\rm coeff}$, $\epsilon_E$, sampling fraction $x\%$, particle-number and spin-projection targets}
\algln{\algkw{Output:} Eigenvalue $E^{(m)}$, CI vector $\mathbf{c}^{(m)}$}
\algln{Prepare $|\Psi(\boldsymbol\theta)\rangle=\hat U(\boldsymbol\theta)|\Phi_0\rangle$ and measure $N_{\rm shots}$ times in the computational basis}
\algln{$\mathcal{B}_{\rm hw}\gets$ sampled configurations with correct particle number and spin projection}
\algln{$\mathcal{B}_{\rm MBPT}\gets$ configurations up to rank $n$ from MP2 amplitudes $t_{ij}^{ab}$ and screened scattering contractions (threshold $\epsilon_{\rm int}$)}
\algln{$\mathcal{B}^{(0)}\gets\mathcal{B}_{\rm hw}\cup\mathcal{B}_{\rm MBPT}$; \ $\mathcal{B}_{\rm blk}^{(-1)}\gets\emptyset$}
\algln{\algkw{for} $m = 0,1,2,\dots$ \algkw{do}}
\algln[1.5]{\textit{Macro-cycle:} solve $\mathbf{H}^{(m)}\mathbf{c}^{(m)}=E^{(m)}\mathbf{c}^{(m)}$ in $\mathcal{B}^{(m)}$}
\algln[1.5]{\algkw{if} $m>0$ \algkw{and} $|E^{(m)}-E^{(m-1)}|<\epsilon_E$ \algkw{then break}}
\algln[1.5]{$\mathcal{B}_{\rm dom}^{(m)}\gets\{|\Phi_\mu\rangle\in\mathcal{B}^{(m)}:|c_\mu^{(m)}|>\epsilon_{\rm coeff}\}$}
\algln[1.5]{$\mathcal{B}_{\rm blk}^{(m)}\gets\mathcal{B}_{\rm blk}^{(m-1)}\cup\{|\Phi_\mu\rangle\in\mathcal{B}^{(m)}:|c_\mu^{(m)}|\le\epsilon_{\rm coeff}\}$}
\algln[1.5]{\textit{Micro-cycle:} train RBM on $\mathcal{B}_{\rm dom}^{(m)}\setminus\{|\Phi_0\rangle\}$ with frequency of $\mu\propto|c_\mu^{(m)}|$}
\algln[1.5]{$\mathcal{B}_{\mathrm{new}}^{(m)}\gets$ symmetry-preserving configurations generated by Gibbs sampling of $x\%$ of the symmetry space}
\algln[1.5]{$\mathcal{B}^{(m+1)}\gets\mathcal{B}_{\mathrm{dom}}^{(m)}\cup\left(\mathcal{B}_{\mathrm{new}}^{(m)}\setminus\mathcal{B}_{\rm blk}^{(m)}\right)$}
\algln[1.5]{\algkw{if} $\mathcal{B}^{(m+1)}=\mathcal{B}^{(m)}$ \algkw{then break}}
\algln{\algkw{return} $E^{(m)}$, $\mathbf{c}^{(m)}$}
\par\vspace{3pt}\hrule height 0.6pt
\end{figure}
% ----------------------------------

Due to the accuracy and efficiency over conventional SQD, in this work, PIGen-SQD is used as a subspace QCSC solver over monomeric fragments.
% efficient fermionic state reconstruction module of PIGen-SQD,
% in this work we use it to be the fragment multi-configurational QCSC solver
% to obtain an accurate description of the effective Hilbert space for each fragment. 
This can be executed in parallel since the fragment eigenstates are
independent of each other in the zeroth order approximation of the theory.
PIGen-SQD would provide $\ket{\Psi_k}$ in terms of a set of associated
coefficients $\{ c^k_{\alpha_k} \}$ and the (local) computational basis $\{ \phi^k_{\alpha_k} \}$ 
for all fragments $k$ under consideration. These set of coefficients and configurations would be used
further to construct the MTPS upon an optimal assignment of computational bit indices, followed by inter-fragment correlation
recovery as discussed in the subsequent sections.

\section{Mapped Aggregation of Partitioned Subspace}

\subsection{Computational Bit Assignment and MTPS Construction} \label{section: computational basis embedding (CBE)}

Our assumption in this work is that 
% Since we assume that
the systems of interest have strong electronic correlations that are local in nature and the associated inter-fragment correlation is relatively weaker. While the local correlations are taken care of by distributed PIGen-SQD, an appropriate assignment
of the local bits to the global indices
is a necessary step to quantitatively reconstruct the $n$-meric configurations from monomeric information.
% Prior to the MTPS construction, appropriate assignment of local computational bits to $n$-meric supersystem bits is necessary for
% appropriate reconstruction of a compact global basis from local basis.
% inter-fragment correlation recovery.
% the improved trainability of the Gen-ML or S-CORE driven inter-fragment correlation recovery.
This assignment provides a direct correspondence between the occupied (1)/unoccupied (0) computational bits of the
physically relevant sub-system configurations and the super-system bits and keeps track of the relative positions
during the reconstruction.
% supersystem and those of its constituent subsystems.
Conceptually, for a fragmentation scheme, an $n$-meric supersystem is partitioned into $n$ disjoint subsystems, $\{S_k\}_{k=1}^n$, each consisting of $M_k$ orbitals and $P_k$ particles. For each subsystem $S_k$, let $\mathcal{I}_k = \{1, 2, \dots, M_k\}$ denote the index set of its local computational bits, where each configuration is represented by a bitstring of length $M_k$.
The global index set of the supersystem is defined as $\mathcal{I}_{\mathrm{g}} = \{1, 2, \dots, M\}$,
where $M = \sum_{k=1}^n M_k$.
The local-to-global index assignment is formalized by a family of injective mappings:
\begin{equation} \label{eq: mapping bit space}
    \pi_k : \mathcal{I}_k \hookrightarrow \mathcal{I}_{\mathrm{g}}, \qquad \forall k \in \{1, \dots, n\}.
\end{equation}
% Since we assume that the systems of interest in this paper have strong electronic correlations that are local in nature and the associated inter-fragment correlation is relatively weaker, such an appropriate assignment is a necessary step to quantitatively reconstruct the $n$-meric configurations from monomeric information.
The proposed assignment consequently maps the local index of each bit to its corresponding $n$-meric index, thereby enabling the systematic construction of the composite multimeric computational basis from the constituent subsystem basis.

For a quantitative description of this assignment problem, we start with the molecular orbitals (MOs) for fragments $\chi_{\beta_k}$ and super-system $\tilde{\chi}_g$, which are linear combinations of atomic orbitals (AOs) assigned to the $k$-th fragment $\xi_{\mu_k}$ and full system $\tilde{\xi}_{\nu}$, respectively:
\begin{equation} \label{eq: Ao to MO expansion}
    \chi_{\beta_k} = \sum_{\mu_k} C_{\mu_k\beta_k} \xi_{\mu_k}, \hspace{5mm} \tilde{\chi}_{g} = \sum_{\nu} \tilde{C}_{\nu g} \tilde{\xi}_{\nu}.
\end{equation}
Here, the notation $\chi_{\beta_k}$ denotes the $\beta$-th MO of the $k$-th molecular fragment, and ${C}$ and $\tilde{C}$ are the associated expansion coefficients for fragments and the super-system.
Since the MOs form the occupation basis in the fermionic Hilbert space, the computational bit assignment
in Eq.~\eqref{eq: mapping bit space} translates to
\begin{equation}
    \pi_k : \chi_{\beta_k} \hookrightarrow \tilde{\chi}_{g}.
\end{equation}
This mapping operation $\pi_k$ computationally translates to a \textit{linear assignment problem} (LAP) and requires the minimization of a cost matrix $J_{g \beta_k}$ for each $g$ and $\beta_k$ defined by (see Appendix \ref{appendix sec: Orbital Mapping and Cost Function} for details)
\begin{equation}
    J_{g \beta_k} = 1 - \left[ |S_{g\beta_k}|^2 \Delta \epsilon_{g \beta_k} \right].
\end{equation}
Here, the cost function $J_{g \beta_k}$ consists of two similarity scores: $S_{g\beta_k}=\sum_{\nu} \sum_{\mu} \tilde{C}_{g \nu} \langle \tilde{\xi}_{\nu} | \xi_{\mu_k} \rangle {C}_{\mu_k \beta_k}$, which is the MO overlap matrix element serving as a metric of spatial similarity, and a Gaussian energy similarity factor $\Delta \epsilon_{g \beta_k} = \exp \left( - \frac{(\epsilon_g - \epsilon_{\beta_k } - \overline{\delta \epsilon_{\beta_k}}) ^2}{2 \sigma ^2} \right)$, where $\epsilon_g$ and $\epsilon_{\beta_k}$ are the orbital energies of the $g$-th global orbital and $\beta$-th local orbital of the $k$-th fragment, $\sigma$ is the energy error tolerance, and $\overline{\delta \epsilon_{\beta_k}}$ is the corresponding mean of the Gaussian distribution (see Appendix \ref{appendix sec: Orbital Mapping and Cost Function}).
The elements of the cost matrix provide a score for each $g$ and $\beta_k$, and our goal is to minimize this cost function such that the local index $\beta_k$ can be optimally assigned a global position $g$. 
This can be solved by the \textit{Hungarian algorithm}~\cite{kuhn1955hungarian}, which is widely used to solve LAPs, particularly in the fields of multi-robot task allocations~\cite{chakraa2023optimization}.
A detailed discussion regarding the LAP and the functional form of the cost function is provided in Appendix~\ref{appendix sec: LAP} and \ref{appendix sec: Orbital Mapping and Cost Function}, respectively.

\textbf{MTPS Construction:} The fragment QCSC solvers (in this case PIGen-SQD) generate a set of dominant configurations for each fragment,
\begin{equation}
\mathcal{B}_k = \left\{ \ket{\phi_1^k}, \ket{\phi_2^k}, \dots, \ket{\phi_{\alpha_k}^k}, \dots, \ket{\phi_{d_k}^k} \right\},
\end{equation}
where $\ket{\phi_{\alpha_k}^k} = \ket{x_{M_k}\dots x_2 x_1} \in \{0,1\}^{M_k}$ represents the computational-basis bitstring for the $\alpha_k$-th configuration of the $k$-th fragment (with individual bits denoted by $x$), and $d_k$ denotes the number of configurations retained for the $k$-th fragment by the fragment PIGen-SQD solver. In practice, configurations with an associated CI coefficient satisfying
\begin{equation}
\left|c_{\alpha_k}^k\right| > \epsilon,
\end{equation}
where $\epsilon$ is a pre-defined threshold, are retained to construct the dominant local configuration space.

Using the computational bit assignment, each local spin-orbital is assigned to its corresponding position in the computational basis of the composite system. The dominant configurations obtained from the individual fragment CI calculations can therefore be embedded into the composite computational basis and combined to construct a compact global subspace. We define this
% set of configurations as
\textit{mapped tensor product state} (MTPS) as
\begin{equation}
\mathcal{B}_{\mathrm{MTPS}} = \{ \ket{\tilde{\Phi}_g} \} = \mathcal{M} \Big( \bigotimes_{k=1}^{n}\mathcal{B}_k \Big),
\label{eq:mapped_product_state}
\end{equation}
where $\ket{\tilde{\Phi}_g} = \ket{\tilde{x}_{M}\dots \tilde{x}_2 \tilde{x}_1} \in \{0,1\}^M$ are the composite super-system configurations
% generated via MTPS
, and $\mathcal{M}(\cdot)$ denotes the mapped tensor-product operation according to the computational bit assignment formalism. 

In contrast to a conventional tensor product, MTPS embeds the local configurations into the computational basis of the composite system according to the orbital assignment.
% established by computational bit mapping.
This distinction is essential because the local configurations $\ket{\phi_{\alpha_k}^k}$ are defined in the Hilbert space of the individual fragments, whereas the resulting configurations $\ket{\tilde{\Phi}_g}$ belong to the Hilbert space of the composite system.
% The computational bit assignment therefore provides the necessary correspondence between local and global spin-orbital indices before the fragment configurations are combined.
The resulting mapped product configurations form a subset of the computational basis of the composite system $\left\{ \ket{\Phi_g} \right\}$,
\begin{equation}
\left\{ \ket{\tilde{\Phi}_{g}} \right\} \subseteq \left\{ \ket{\Phi_g} \right\},
\end{equation}
and define a compact subspace of the exponentially large Hilbert space. Since this subspace is constructed from the dominant configurations of the individual fragment CI solutions, it is expected to retain substantial support to the target eigenstate of the full Hamiltonian. The resulting compact subspace can subsequently be used for projection and diagonalization of the full Hamiltonian, thereby providing an efficient route to incorporate inter-fragment correlation beyond the zeroth-order product-state description.

\textbf{Cross-Fragment Configurations from Local CI Solutions:}
The MTPS constructed from the individual monomer determinants intrinsically accounts for intra-fragment excitations and embeds them appropriately in the Fock space of the composite system. However, in general, it does not contain inter-fragment correlation effects, which include configurations corresponding to electronic excitations between the constituent fragments. Such inter-fragment excitations can be particularly important for weakly bound donor-acceptor complexes, where the frontier orbitals of the composite system may be localized predominantly on different monomeric subunits. The omission of these configurations can therefore restrict the description of inter-fragment charge redistribution and charge-transfer characters. To account for these effects, we augment the mapped product space with a set of cross-fragment excited configurations.
% constructed within a pre-defined donor-acceptor orbital window.

Since the origin of each orbital (or bit in the computational basis) is identified and can be associated with a particular fragment using the computational bit assignment, cross-fragment terms can be generated by exciting electrons from each determinant of a particular fragment to the unoccupied spinorbitals of neighboring fragments. Although the number of all cross-fragment excited configurations grows combinatorially with the number of electrons in one fragment and the number of unoccupied orbitals in the other fragments, we restrict this expansion by applying a one-body cross-fragment excitation (within the same spin sector) from $\ket{\phi_{\alpha_k}^k}$ to $\ket{\phi_{\alpha_l}^l}$ (and vice versa) only if the numerical product of the corresponding monomeric CI coefficients satisfies
\begin{equation} \label{eq: cross-fragment MTPS condition}
\left|c_{\alpha_k}^{k}c_{\alpha_l}^{l}\right| > \epsilon_{cf},
\end{equation}
where $c_{\alpha_k}^{k}$ and $c_{\alpha_l}^{l}$ denote the corresponding coefficients of the associated determinants, and $\epsilon_{cf}$ is the cross-fragment threshold.
This set of cross-fragment configurations supplements the mapped product space and explicitly introduces low-rank inter-fragment correlation effects.
From here onwards, the definition of MTPS includes collectively Eq. \eqref{eq:mapped_product_state} as well
as the cross fragment single excitations discussed here.

\begin{figure*}[!ht]
    \centering  
\includegraphics[width=\textwidth]{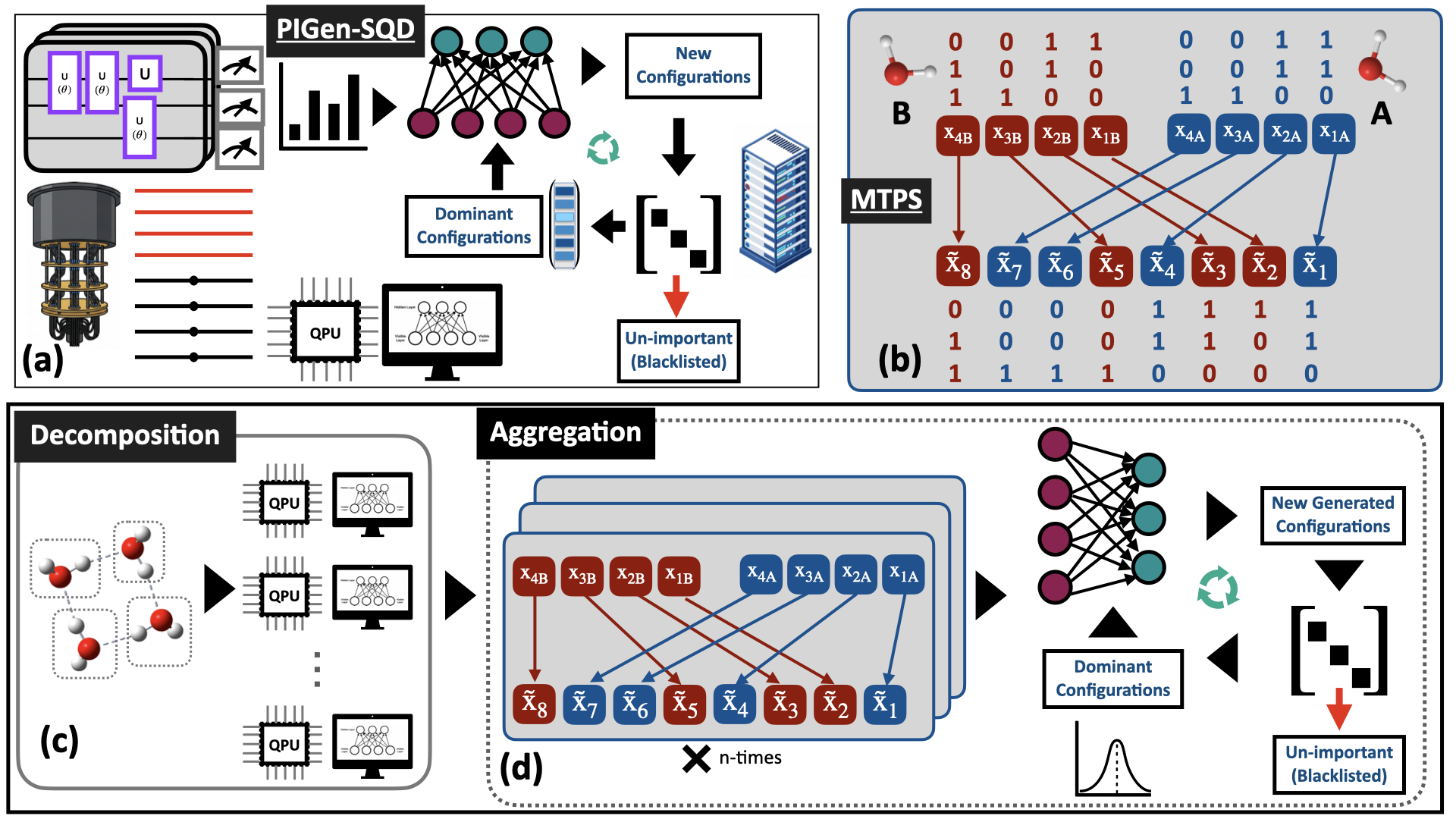}
\caption{(a) PIGen-SQD workflow: A molecular system is encoded into a PQC and measurement is done to obtain the samples. These samples are then
classically post-processed with Gen-ML driven modules for fermionic configuration reconstruction. (b) Computational bit mapping and MTPS construction: The orbitals of two sub-systems, A and B are appropriately assigned to the super-system bits.
The dominant configurations obtained by distributed PIGen-SQD is then translated to the global bits from local bits
accordingly. PIGen-SQD(MAPS) workflow is shown in the lower panel with (c) showing the problem decomposition and distributed PIGen-SQD solver, followed by
(d) aggregation with MTPS and Gen-ML driven inter-fragment correlation recovery.
}
    \label{fig: algorithmic workflow schematics}
\end{figure*}

\subsection{Systematic Recovery of Inter-Fragment Correlation Effects} \label{sec:inter-frag corr recovery}

The MTPS construction described in Section \ref{section: computational basis embedding (CBE)} exploits the chemically motivated fragmentation involving
strong local electronic correlations and partially recovers the inter-fragment correlation effects with low-rank cross-fragment excitations.
% as already discussed
% in section \ref{section: computational basis embedding (CBE)}.
While MTPS provides a substantial overlap with the target state, a systematic recovery of inter-fragment correlation is necessary for
an accurate description of the composite system. Toward this, MTPS provides an
extremely good initialization for subsequent self-consistent inter-fragment correlation recovery.
% \textcolor{blue}{In this workflow, two different types of modules are considered for such recovery: one at the fragment QCSC solver level
% (discussed in section \ref{subsec: pigen-sqd})
% and one at the inter-fragment recovery level.
% and in general one may use any algorithm for the same.
% Here we have considered a secondary layer of Gen-ML model for inter-fragment recovery
% which is basically the same module used in PIGen-SQD.
% Additionally, we have studied S-CORE, which is used in conventional SQD, initialized by MTPS as well (to be called S-CORE(MTPS))
% to explicitly see the effect of MTPS as an initialization to configuration recovery modules compared to
% standard SQD.
% and to be called S-CORE initialized by
% MTPS or S-CORE(MTPS), and a
% Gen-ML driven recovery which is used in standard PIGen-SQD workflow.
% In this sense
% So, overall in this work we are using PIGen-SQD as our primary QCSC solver and we will demonstrate the performance
Here we consider two possible inter-fragment correlation recovery methods as follows:
% one may use any fragment QCSC solver followed by inter-fragment correlation recovery
% PIGen-SQD is the primary fragment QCSC solver are used,
% and one may subsequently use any 
% another one at inter-fragment recovery level, we refer to this algorithm as PIGen-MTPS-SQD($\mathrm{X}$),
% where $\mathrm{X}$ is the corresponding inter-fragment recovery module.}
% Here we discuss two possible inter-fragment correlation recovery methods in the QCSC framework as follows:

% \subsubsection*{1. Self-Consistent Configuration Recovery Initialized by MTPS}

\textbf{S-CORE Initialized by MTPS: }The self-consistent configuration recovery or S-CORE module\cite{robledo2025chemistry} in the SQD protocol provides a systematic recovery of dominant determinants for accurate electronic correlation
energy estimation with Hamiltonian projection and diagonalization. However, the accuracy and efficiency of S-CORE depends on the initial point of the iterative procedure
in a way that a better initial set of dominant configurations leads to more accurate and more compact diagonalization subspace.
Thus for extended systems with multiple chemically motivated fragments, MTPS qualifies as a natural choice for such an initialization. In this paper we
have used S-CORE with \textit{carryover mechanism}\cite{wang2026localized}, which uses an additional retention step to save and carry important set of bitstrings from one iteration
to the next iteration to ensure that energetically relevant configurations are incorporated in the recovery process.
This inter-fragment recovery with S-CORE initialized by MTPS will be referred to as S-CORE(MTPS). 
% \textcolor{blue}{It is important to mention here that, since the efficiency and accuracy of the iterative inter-fragment correlation recovery requires a fairly accurate initialization. Due to this reason, for monomers, we have used PIGen-SQD as it has been demonstrated\cite{Patra_2026} that it provides more dominant and more compact configuration space which is essential for MTPS construction.} 
% \textcolor{red}{ALSO MENTION THAT FOR THE MONOMERIC PROBLEMS, WE STILL REQUIRED/USED PIGen-TYPE CR AND S-CORE DEFINED HERE ARE ONLY FOR THE DIMERIC PROBLEMS.}

% \subsubsection*{2. Generative Machine-Learning-Driven Recovery}

\textbf{Generative Machine-Learning-Driven Recovery: }
A more sophisticated and systematic approach would be to self-consistently train a secondary layer of Gen-ML model initialized with MTPS to search and select unseen dominant sectors of the Hilbert space.
So in this entire workflow, we use two layer of Gen-ML: one in PIGen-SQD and one in the cross-fragment recovery stage
and both the stages use the same RBM module discussed in details in Appendix \ref{appendix: RBM}.
This pipeline consisting of PIGen-SQD, MTPS and Gen-ML driven cross-fragment recovery will be referred to
as PIGen-SQD \textit{with mapped aggregation of partitioned subspace} or PIGen-SQD(MAPS).

To summarize,
the PIGen-SQD(MAPS) algorithm involves the following steps:
\begin{enumerate}
    \item \textbf{Parallel Fragment QSCI Solvers:} The molecular system is divided into chemically motivated fragments with dominant localized electronic correlations. Once the optimal fragments are obtained, PIGen-SQD (as discussed in section \ref{subsec: pigen-sqd}) modules are used for each fragment as distributed QCSC solver.

    \item \textbf{Computational Bit Assignment and MTPS Construction: } After the convergence of PIGen-SQD cycles, all the dominant configurations based on their associated coefficients are collected. The computational bit positions are appropriately assigned according to the approach described in Section \ref{section: computational basis embedding (CBE)} to map each local bit to a global bit. This enables one to associate local orbitals to global orbitals.
    % which is essential for subsequent MTPS construction.
    With the assigned bits, MTPS is constructed from dominant fragment configurations obtained from PIGen-SQD.
    % as described in Section \ref{subsec: pigen-sqd}.
%     For an extended system if one uses FMO-n approximation then the MTPS is restricted \textcolor{red}{up} to
% \begin{equation}
% % \left\{ \ket{\Phi_{\alpha_1,\ldots,\alpha_{n}}} \right\} = \mathcal{M} \Big( \bigotimes_{k=1}^{n}\mathcal{B}_k  \Big)
% \mathcal{B}_{\mathrm{MTPS}} = \{ \ket{\tilde{\Phi}_g} \} = \mathcal{M} \Big( \bigotimes_{k=1}^{n}\mathcal{B}_k  \Big)
% %\left\{ \ket{\phi_{\alpha_1}^1} \right\} \overset{\mathcal{M}}{\otimes} \left\{ \ket{\phi_{\alpha_2}^2} \right\} \overset{\mathcal{M}}{\otimes} \cdots \overset{\mathcal{M}}{\otimes} \left\{ \ket{\phi_{\alpha_{N_f}}^{N_f}} \right\},
% \label{eq:mapped_product_state fmo variant}
% \end{equation}
% % Such formulation of MTPS ought to be performed for all possible n-meric combinations and all possible $(n-1), (n-2), (n-3), ...2$-meric calculations.

    \item \textbf{Systematic Recovery of Inter-fragment Correlation:} Inter-fragment correlation, which is only partially recovered MTPS construction, is self-consistently recovered using standard S-CORE initialized with MTPS or a secondary layer of Gen-ML module, as discussed in Section \ref{sec:inter-frag corr recovery}. 
\end{enumerate}
The associated schematics of the workflow is shown in Fig. \ref{fig: algorithmic workflow schematics}.

Most importantly, in this workflow, we have entirely circumvented the circuit preparation for the composite system, which would otherwise require substantially deeper circuits and extremely high number of shots for meaningful sampling. The mapped product state is directly fed into the S-CORE or Gen-ML based configuration recovery module, which self-consistently recovers additional dominant configurations until convergence for an accurate description of the
composite system while only exploring a restricted space at a time.

\section{Results and Discussion}

\begin{table*}[t]
\caption{\textbf{Details of number of spinorbitals, number of qubits (including auxiliary qubits), number of shots, FCI dimension, two-qubit quantum gate count, and two qubit circuit depth
for the molecular systems studied with the LUCJ ansatz.}}
\label{tab:quantum_resources}
\centering
% Adjust this length to change the spacing between columns (default is usually 6pt)
\setlength{\tabcolsep}{6pt} 
\begin{tabular*}{\textwidth}{@{\extracolsep{\fill}}lcccccc}
\toprule
System (Active Space) & 
Spinorbitals & 
Qubits & 
$N_{\text{shots}}$ & 
$\dim(\text{FCI})$ & 
Gate Count & 
Gate Depth \\
\midrule
$\text{H}_2\text{O}$ (8e, 10o)       & 20   & 23   & $2 \times 10^5$ & 44100 & 524        & 86   \\
$\text{(H}_2\text{O)}_2$ (16e, 20o)   & 40   & 45   & $5 \times 10^6$ & $1.58 \times 10^{10}$ & 2072       & 154  \\
$\text{FMA}$ (18e, 12o)       & 24   & 27   & $2 \times 10^5$ & 48400 & 644        & 82   \\
$\text{(FMA)}_2$ (36e, 24o)   & 48   & 54   & $3 \times 10^6$ & $1.81 \times 10^{10}$ & 2672       & 170  \\
\bottomrule
\end{tabular*}
\end{table*}

\subsection{Computational details}
\textbf{Molecular Calculations and Fragmentation:} {In this paper, biochemically relevant $\mathrm{(H_2O)_2}$ and formamide (FMA) dimer (geometry given in Supplementary Information)
are studied at different intermolecular separations.
In particular, accurate electronic structure calculations of FMA dimers are of immense biochemical importance as it is the smallest system containing the peptide bond.
For the dimeric systems (which is the
full system in this particular example), restricted Hartree-Fock (RHF) calculations are performed with PySCF using 6-31G basis set. The dimer active spaces for $\mathrm{(H_2O)_2}$ and FMA homo-dimer are chosen
to be (16e, 20o) and (36e, 24o) which forms $\mathrm{1.58 \times 10^{10}}$ and $\mathrm{1.81 \times 10^{10}}$ dimensional FCI space respectively.
% was obtained by removing lowest 6 core spatial orbitals for formamide dimer and 4 core spatial orbitals for water dimer along with $!!!!$ and $!!!!$ highest spatial orbitals for formamide and water dimer respectively.
Effective one-electron and two-electron active space integrals with the constant core energy contributions are obtained from PySCF CASCI integral transformations.} 
For distributed QCSC calculations, the dimeric systems are fragmented
into two monomeric sub-systems such that each fragment of $\mathrm{H_2O}$ and FMA consists of (8e,10o) and (18e,12o) active spaces
with $\binom{10}{4}^2=$ 44100 and $\binom{12}{9}^2=$ 48400 dimensional FCI space respectively.
Calculations with GAMESS-US\cite{schmidt1993general} are performed to obtain fragment-resolved
molecular orbital coefficients, one-electron integrals, two electron integrals, fock matrices, density matrices and nuclear-repulsion energies for each monomer, embedded in the mean-field of the other monomer.
Also, to test the capabilities of the PIGen-SQD(MAPS) workflow for studying extended systems, we have studied $\mathrm{(H_2O)}_6$ (geometry given in Supplementary Information)
system under the FMO-2 approximation.
% In this case, all the necessary monomeric and dimeric
% integrals are obtained with the mean-field effect of the other parts of the system using GAMESS.
Different classical SCI methods have been studied such as Heat-bath CI (HCI) and
standard SCI implementation in PySCF (which we will refer to as SCI-PySCF)
for comparing the performance of PIGen-SQD(MAPS).
HCI results are obtained with \texttt{qiskit\_addon\_dice}\cite{sharma2017semistochastic, holmes2016heat} package
with the variational selection threshold kept at $10^{-5}$,
while SCI-PySCF results are obtained with \texttt{pyscf.fci.selected\_ci} routine in
PySCF\cite{sun2018pyscf}.
% \textsc{pyscf}~\cite{sun2018pyscf}.
% The wave function expansion was controlled 
% using a selection threshold of $\epsilon = 10^{-4}$

\begin{figure*}[!ht]
    \centering  
\includegraphics[width=\textwidth]{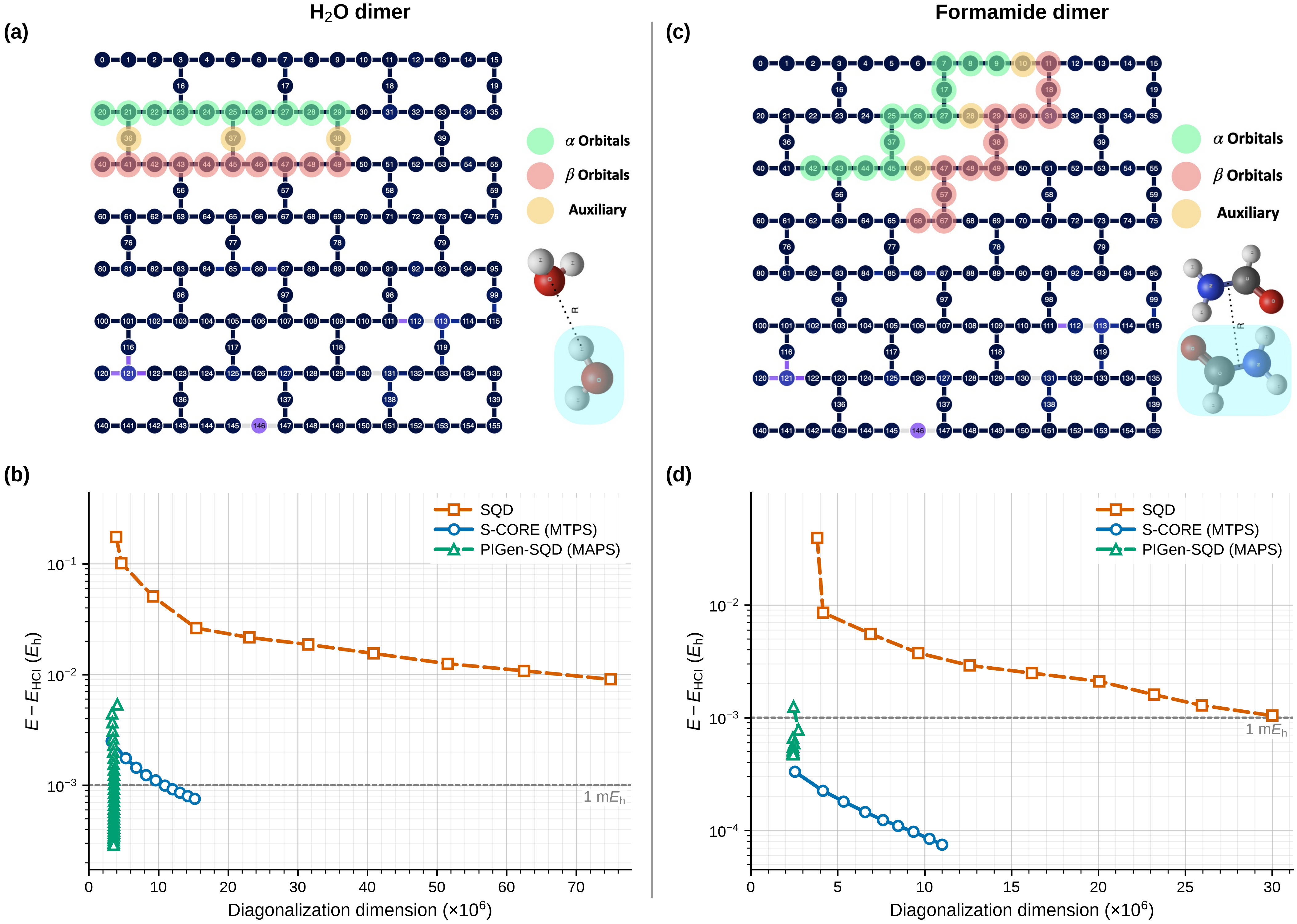}
\caption{
Coupling map and qubit assignment on IBM Heron R3 processor for monomeric fragments of water and FMA dimer are shown in (a) and (c) respectively. The intermolecular Oxygen-Oxygen
distance between two $\mathrm{H_2O}$
monomers is 2.2 {\AA} and the perpendicular distance between the parallel C--N bonds of the two FMA monomers is 3.23 {\AA}. Here $\alpha$ ($\beta$) orbitals are shown in green (red) along with the auxiliary qubits (in yellow) that encode density-density interactions in the LUCJ ansatz. Energy error (in $\mathrm{E_h}$) with respect to HCI and the associated diagonalization dimension are
shown in (b) and (d) for SQD, S-CORE(MTPS) and PIGen-SQD(MAPS). Grey dotted line shows the 1 $\mathrm{mE_h}$ error with respect to HCI.
% \textcolor{red}{THE CORRESPONDING DISCUSSION IN THE TEXT MUST STRESS THE IMPORTANCE OF CBE AND MTPS FORMULATION FOR S-CORE(MTPS) AND PIGEN(MTPS) AS ANY FAULTY ASSIGNMENT OF THE LOCAL ORBITALS TO THE CANONICAL BASIS MEANS EXTREMELY INCREASED WORKLOAD FOR THE CR MODULE AND PRESUMABLY THIS MAY LEAD TO THE INFLATION OF THE DIMENSION WITHOUT ACCOUNTING ANY CORRELATION. \\~\\ MENTION THAT THESE ARE THE DIMENSIONS FOR THE \textbf{DIMERIC} PROBLEMS POST S-CORE or PIGen type CR BASED ON MTPS VS FROM DIMERIC SAMPLING}}
}
    \label{fig: coupling map and S-CORE with MTPS}
\end{figure*}

\textbf{Quantum Circuit Details:} In the PIGen-SQD(MAPS) workflow, PIGen-SQD routines are used in parallel for each monomer which requires monomeric quantum
hardware samples followed by Gen-ML driven fermionic state reconstruction\cite{Patra_2026}.
The perturbative configurations in the PIGen-SQD workflow are obtained with sparse-tensor operations and the corresponding algebraic terms are given in Appendix \ref{appendix: PT algebraic terms}. 
Conventional
SQD is also studied over the dimeric systems to compare the results that uses
the samples obtained for dimer system followed by S-CORE modules with \textit{carryover} mechanism\cite{wang2026localized}
as implemented in \texttt{qiskit\_addon\_sqd}\cite{qiskit-addon-sqd} (version 0.12.1) with carryover threshold set at $10^{-6}$.
LUCJ ansatz with optimized CCSD parameters are
implemented with \texttt{ffsim}\cite{ffsim} package for both monomeric and dimeric systems. The corresponding spinorbitals are mapped onto \texttt{ibm\_boston}
hardware that runs on IBM Heron R3 processor with heavy-hex coupling map.
The \texttt{qiskit\_zigzag\_layout} function is used for the qubit assignments on quantum processors.
S-CORE(MTPS) results are obtained with standard carryover S-CORE module.
The hyperparameters used for SQD and S-CORE(MTPS) are given in Table \ref{tab:SQD_hyperparameters} in Appendix \ref{app: hyperparameters for SQD and S-CORE(MTPS)}. 
For PIGen-SQD(MAPS) workflow, only monomeric hardware samples are used along with multi-layered Gen-ML driven recovery modules to get the
solutions for each fragment and for subsequent inter-fragment recovery.
No dimeric hardware samples are used for PIGen-SQD(MAPS) framework to demonstrate it as a method with low shot requirement.
However, it is straightforward to add dimeric samples as well along with MTPS.
% In this case, no dimeric hardware samples are used to
% show the advantage of MTPS construction followed by configuration recovery with limited shots over conventional SQD performed over the dimeric system that requires quantum hardware samples for the dimers.
The coupling map of the QPU and associated qubit assignments are shown in Fig. \ref{fig: coupling map and S-CORE with MTPS} (a, c)
for $\mathrm{H_2O}$ and FMA monomer. The details regarding the system active space, number of spinorbitals, number of qubits (including the auxiliary qubits),
$N_{\mathrm{shots}}$, FCI dimension, two-qubit gate count and two-qubit gate depth are shown in Table \ref{tab:quantum_resources}.

% while that of the
% dimers are shown in \textcolor{blue}{ADD SUPPL ALSO ADD THE CIRCUIT DETAILS: DEPTH, CNOT COUNT ETC.}

% initial subspace is constructed with the HARDWARE DATA obtained from the IBM QUANTUM COMPUTER, followed by a generative ML based configuration recovery (GenML-CR) with Restricted Boltzmann Machine (RBM).
% After initial determinant-space (hardware samples) diagonalization, determinants are partitioned into dominant and blacklisted subsets according to the magnitude of their CI coefficients.
% In the numerical calculations
% reported in this work,

\begin{figure}
    \centering
    \includegraphics[width=1.0\linewidth]{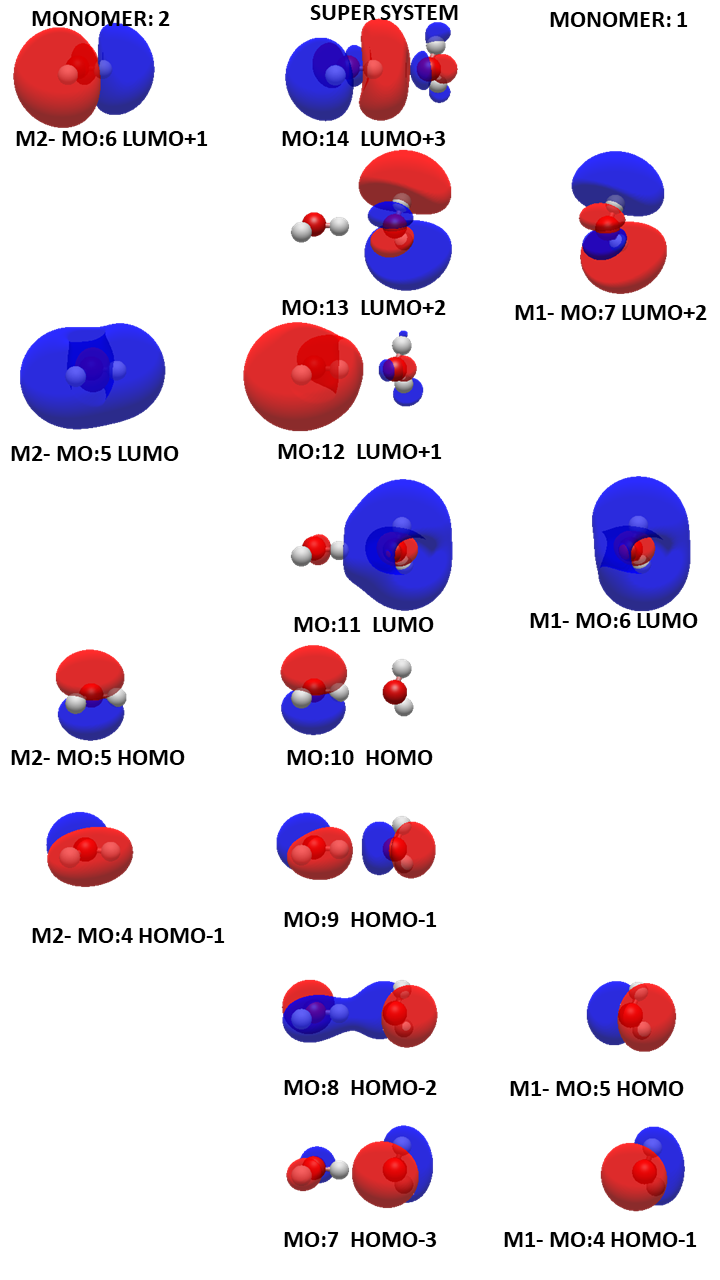}
    \caption{Orbital correspondence in $\mathrm{(H_2O)_2}$. (Center) Subset of dimer molecular orbitals situated near the HOMO-LUMO region. (Right, Left) Assigned monomeric orbitals for monomer 1 (M1) and monomer 2 (M2), respectively, established through computational bit assignment.
    % Occupied and virtual orbitals of monomer-1 (right) and monomer-2 (left), lying close to the HOMO--LUMO window, are shown together with their corresponding supersystem orbitals (center).(\textcolor{red!65!black}{need horizontal lines to show one to one assignments}. Fragment orbitals (optimized in the field of external potential created by the surrounding fragment.) are embedded in the supersystem atomic-orbital basis, and the assignment cost combines the squared AO-metric orbital overlap with a Gaussian orbital-energy compatibility factor.) The orbitals are shown with same isosurface value and molecular orientations. Red and blue colour denotes the global phase of the orbitals; the overall phase is arbitrary and does not affect the mapping because square of the orbital orbital ($|O_{ij}|^2$) is used in the assignment.
    }
    \label{fig:computational basis mapping}
\end{figure}

% \textbf{Computational Basis Mapping and MTPS:} For computational basis mapping, the associated LAP (as described in Section \ref{section: computational basis embedding (CBE)})
% was solved using the
% \texttt{scipy.optimize.linear\_sum\_assignment} module as implemented in
% SciPy version~1.17.0.\cite{Virtanen2020}. The routine implements Crouse's\cite{Crouse2016} variant which is one of the modified
% and efficient variants of the original Hungarian algorithm among several existing other implementations\cite{JonkerVolgenant1986, JonkerVolgenant1987}.

\textbf{Computational Bit Assignment, MTPS and Inter-fragment Correlation Recovery:} The LAP associated with computational bit assignment (Section ~\ref{section: computational basis embedding (CBE)}) was solved via the \texttt{scipy.optimize.linear\_sum\_assignment} routine in SciPy (v1.17.0)~\cite{Virtanen2020}. The routine is based on Crouse's formulation~\cite{Crouse2016}, representing one of several highly efficient adaptations of the original Hungarian algorithm~\cite{JonkerVolgenant1986, JonkerVolgenant1987}.

% which was developed for better efficiency.
% Jonker-Volgenant (JV) algorithm\cite{JonkerVolgenant1986, JonkerVolgenant1987}.
% JV algorithm was introduced as an attempt to improve the performance of the original Hungarian algorithm
% and they solve the same linear assignment
% problem and minimize the same cost matrix; they differ only in the numerical
% procedure used to obtain the optimal one-to-one assignment.
% The SciPy module uses explicitly the implementation described by Crouse\cite{Crouse2016}.
% algorithm without a separate initialization stage, following the implementation described by Crouse.\cite{Crouse2016}.
% The Hungarian and JV algorithms solve the same linear assignment
% problem and minimize the same cost matrix; they differ only in the numerical
% procedure used to obtain the optimal one-to-one assignment.

% The Jonker Volgenant algorithm \cite{JonkerVolgenant1987}, which was introduced as an attempt to improve of the hungarian assignment problem \cite{JonkerVolgenant1986}

\begin{figure}
    \centering
    \includegraphics[width=1.0\linewidth]{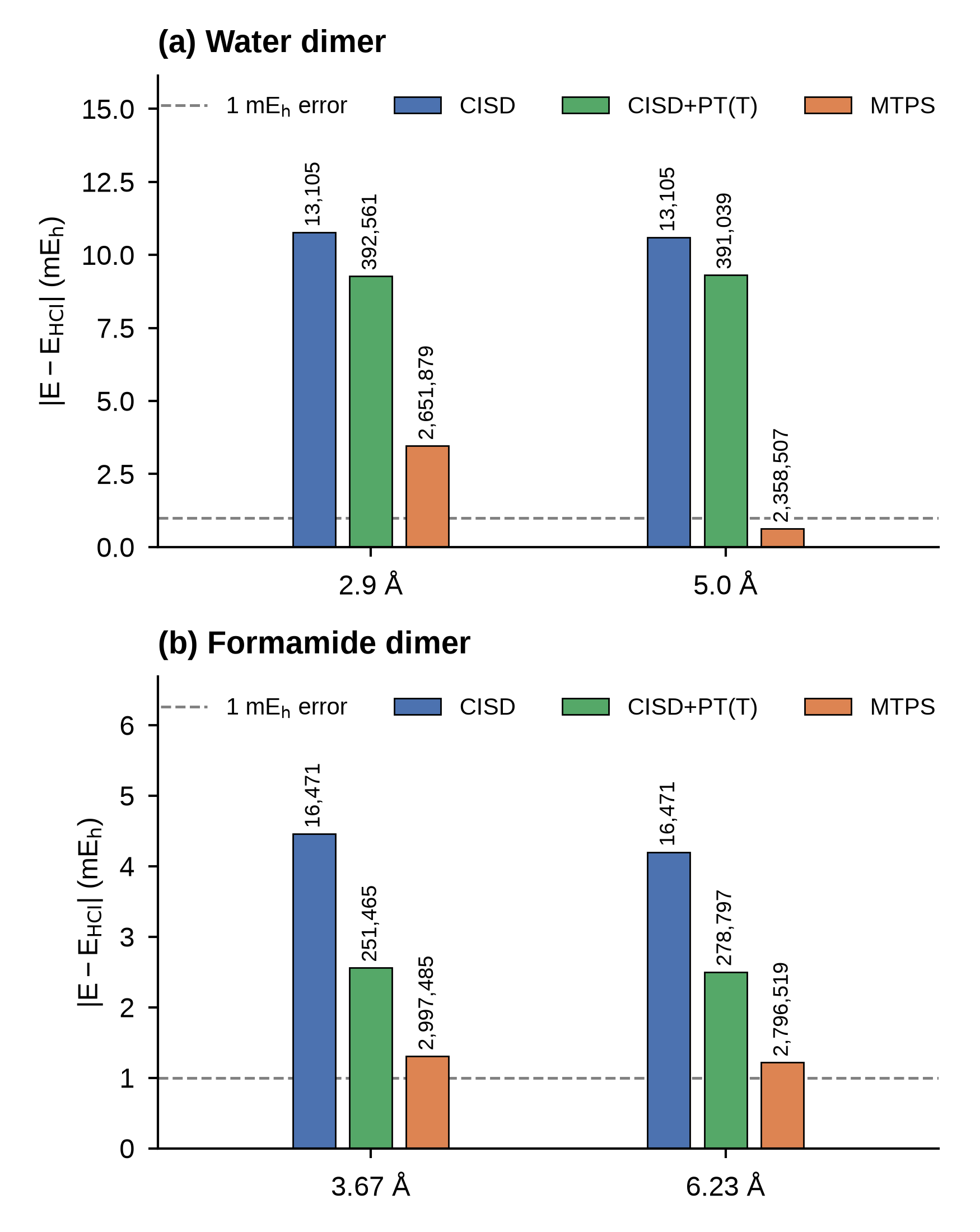}
    \caption{Contribution of CISD, CISD + perturbative triples and MTPS configurations toward recovery of correlation energy for water and FMA dimer for two different separation distances each.
    In y-axes energy error (in $\mathrm{mE_h}$) with respect to HCI is shown.
    On top of each bar plot, the associated diagonalization dimension is also shown. The dotted line shows 1 $\mathrm{mE_h}$ error from HCI.}
    \label{fig:mtps energy bar plot}
\end{figure}

The associated MTPS construction is done in two phases which incorporate both intra and inter-fragment correlation effects as described in Section \ref{section: computational basis embedding (CBE)}.
For generating the intra-fragment contributions, the dominant
configurations from each monomer are assembled based on a pre-defined cut-off threshold
$\left|c_{\alpha_k}^k\right| > \epsilon$ (in this case  $\epsilon =10^{-6}$.
% The coefficient cutoff for monomers were
% The dominant determinants with coefficients higher than the cutoff are retained for fragment-product space construction.
% Second step of our protocol is to map the fragment monomer orbitals to the canonical orbital basis using LAP discussed in the section II B. Canonical product determinants (PDT) were constructed from mapped products of the converged monomer determinant expansions.
With the assumption that strong correlation effects have intra-fragment characteristics, and to restrict the dimensionality of the product space,
% Before taking product dominant
the monomer configurations were ordered by decreasing coefficient magnitude and
MTPS is formed only with those $c_{\alpha_k}^k$ and $c_{\alpha_l}^l$ that satisfy
$\left|c_{\alpha_k}^k c_{\alpha_l}^l\right| > 10^{-7}$.
% product determinant formed from monomer states I and J was retained when
% \begin{equation}
% \left|c_{\alpha}^I c_{\beta}^J\right| > 10^{-7}
% \end{equation} 
The corresponding occupation vectors were then permuted into the canonical dimer orbital ordering using the LAP. Product-space generation was parallelized over all available CPU cores.
For generating the inter-fragment sector of the MTPS we have used $\epsilon_{cf}=10^{-5}$ in Eq. \eqref{eq: cross-fragment MTPS condition}.
Inter-fragment excitations are generated in a way such that the $\hat{S}_z$ spin symmetry of the super-system is maintained.
This cross-fragment excitation generation is parallelized similar to the mapped product space generation and duplicate determinants were removed.
HF, CISD and perturbatively obtained triples (see Appendix \ref{appendix: PT algebraic terms}) for dimeric system
are augmented with MTPS to provide a better initialization for the subsequent configuration recovery modules.
Spin-mirror partners, generated by interchanging the up and down spin occupation sectors, were subsequently included.
This is followed by a final removal of any additional duplicate determinants in the entire set of configurations
% This combined set of configurations may contain duplicate elements which are subsequently removed
and this de-duplicated set of configurations are passed to the diagonalizer.
% \textcolor{red}{In this case we have used a modified variant of
% the DICE solver from \texttt{qiskit\_addon\_dice}}.
Note that from here onwards, MTPS refers to this combined subspace which contains manual augmentation of HF, CISD
perturbative triples followed by de-duplication to ensure that no obvious lower order dominant configurations are left out.
In the next section (Section \ref{subsec: result water and formamide dimer}), the numerical demonstrations different inter-fragment recovery
mechanisms initialized with MTPS are discussed in details.

\subsection{Water and Formamide Dimer} \label{subsec: result water and formamide dimer}

\textbf{Computational Bit Assignment:} One of the most important steps of the PIGen-SQD(MAPS) workflow is the computational bit assignment from local monomeric system to
$n$-meric system since an optimal local to global mapping is essential for accurate reconstruction of supersystem
orbitals from monomeric PIGen-SQD solutions. 
Fig. \ref{fig:computational basis mapping} shows a subset of dimer orbitals for $\mathrm{(H_2O)_2}$ adjacent to highest occupied MO (HOMO)
and lowest unoccupied MO (LUMO) region (center) along with the assigned monomeric orbitals, denoted as
M1 (right) and M2 (left) for monomer 1 and 2, respectively, obtained with computational bit mapping.
% Pictorial representation of fragments to Supersystem correspondence map for the water dimer.
% Occupied and virtual orbitals of M1 (right) and M2 (left), lying close to the HOMO--LUMO window, are shown together with their corresponding supersystem orbitals (center).
% Fragment orbitals (optimized in the field of external potential created by the surrounding fragment) are embedded in the supersystem atomic-orbital basis, and
The assignment cost function combines the squared AO-metric orbital overlap with a Gaussian orbital-energy compatibility factor as discussed in section
\ref{section: computational basis embedding (CBE)}. The orbitals are shown with the same
isosurface value and molecular orientations. Red and blue colour denote
the global phase of the orbitals, however, the overall phase is arbitrary and does not affect the mapping because square of the orbital overlap
($|S_{g\beta_k}|^2$) is used in the assignment.
Fig. \ref{fig:computational basis mapping} pictorially represents that the
LAP used in this study is fairly optimal as the local molecular orbitals
retain their shape in the dimeric super-system (up to a constant phase factor).
The orbital assignment for FMA dimer is shown in Fig. \ref{fig:computational basis mapping FMA dimer}
in Appendix \ref{appendix: formamide mapped orbitals}.

\begin{figure*}[!ht]
    \centering  
\includegraphics[width=\textwidth]{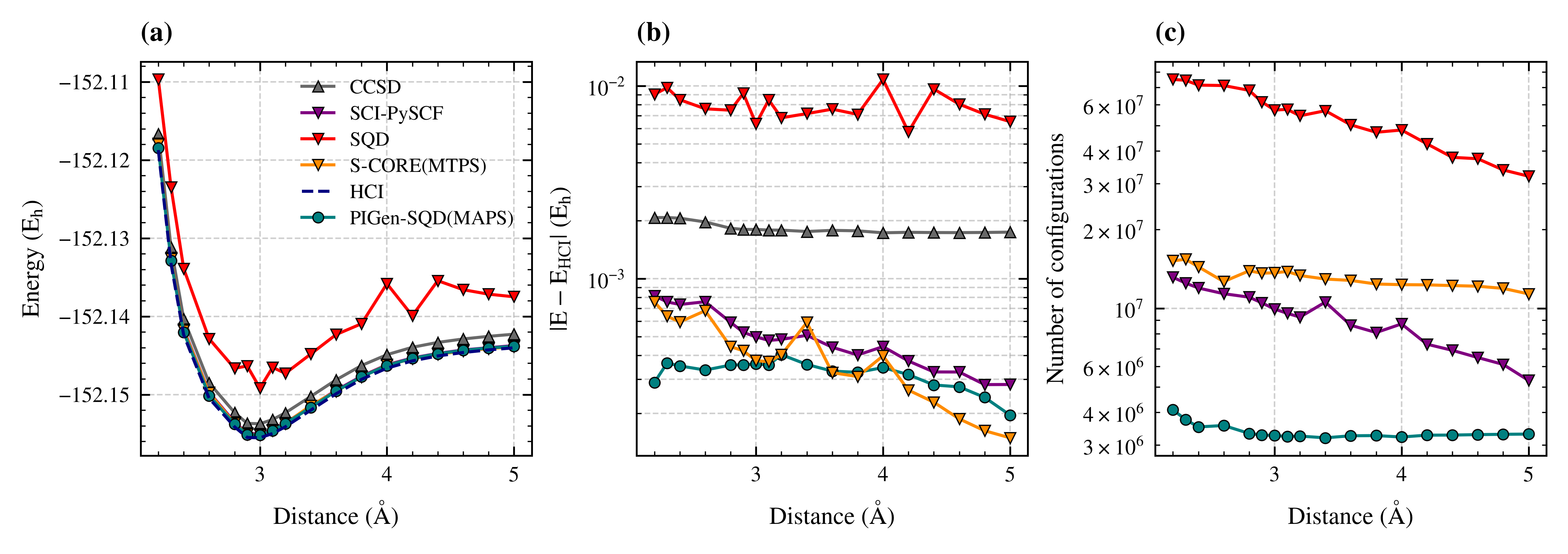}
\caption{Potential energy surface (PES) for $\mathrm{(H_2O)_2}$. Along the y-axes (a) shows the energy (in $\mathrm{E_h}$) with different methods: CCSD, SCI-PySCF, SQD, S-CORE(MTPS), PIGen-SQD(MAPS) and HCI, (b) shows error with respect to HCI and (c) shows the corresponding diagonalization dimension. Different Oxygen-Oxygen distance (in \AA) between two monomers are represented along x-axes.}
    \label{fig: pes plot H2O dimer}
\end{figure*}

\begin{figure*}[!ht]
    \centering  
\includegraphics[width=\textwidth]{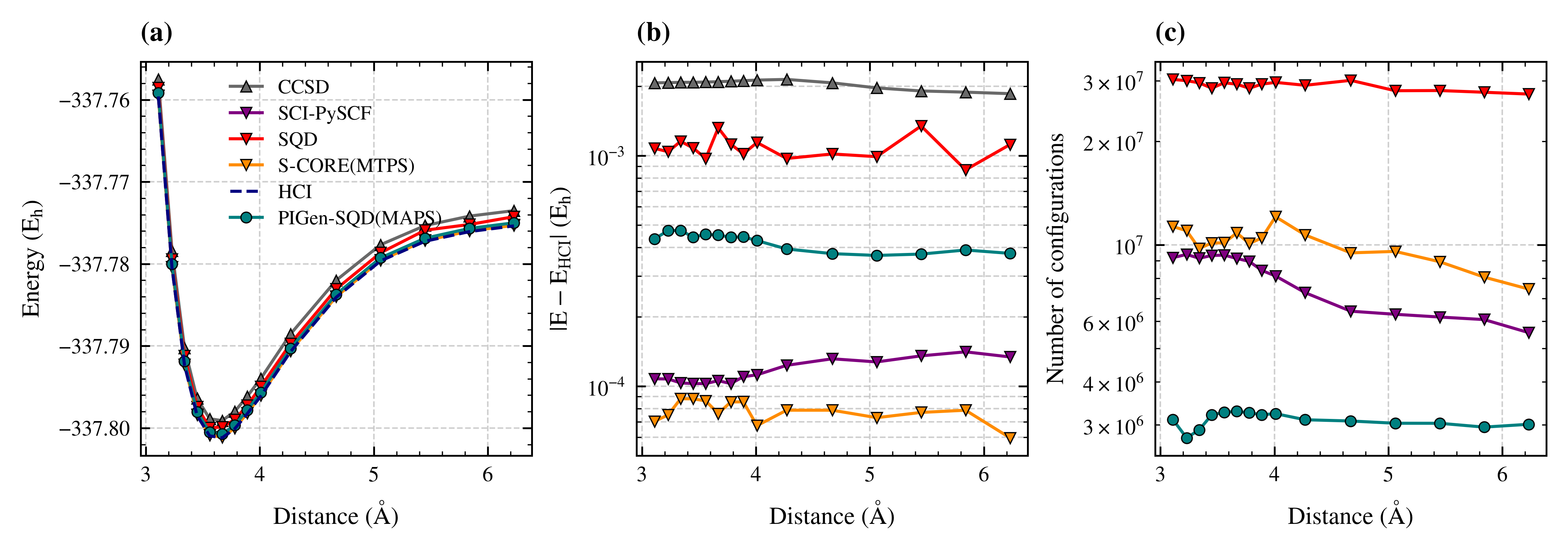}
\caption{PES and other quantities for $\mathrm{(FMA)_2}$. Axes information are the same as that of Fig. \ref{fig: pes plot H2O dimer}. }
    \label{fig: pes plot formamide dimer}
\end{figure*}

\textbf{MTPS:} Fig. \ref{fig:mtps energy bar plot} shows the contribution of MTPS toward the recovery of electronic
correlation directly from monomeric solutions without any explicit iterative correlation recovery performed over dimeric system.
% Here, MTPS configurations are augmented with HF, CISD and dominant triples configurations obtained perturbatively\cite{Patra_2026}
% and the energy is obtained upon diagonalization in this subspace.
The y-axes show the energy error with respect to HCI (in $\mathrm{mE_h}$)
and on top of each bar plot, the associated diagonalization dimension is mentioned.
It shows the systematic improvements upon incorporating the perturbative triples and MTPS
over CISD determinants. For equilibrium geometries, MTPS shows around 7 $\mathrm{mE_h}$ and 3 $\mathrm{mE_h}$ 
improvements over CISD for $\mathrm{(H_2O)_2}$ and $\mathrm{(FMA)_2}$ respectively.
For stretched inter-molecular separation of water dimer, MTPS
is extremely accurate (< 1 $\mathrm{mE_h}$ error)
since electronic correlations effects are highly localized and thus MTPS provides an accurate description of the weak
inter-molecular interactions in such cases.

\textbf{Inter-Fragment Correlation Recovery:} Since the performance of any inter-fragment correlation recovery module, be it Gen-ML or S-CORE driven, is sensitive to the
quality of the initial seed, MTPS provides extremely good initialization for the same.
To demonstrate this, in Fig. \ref{fig: coupling map and S-CORE with MTPS} various configuration recovery
strategies are compared.
For a fair and head-to-head comparison, a study is shown between standard SQD with quantum hardware samples seed (obtained from
circuits constructed for dimeric systems) and
S-CORE(MTPS), which refers to
the standard S-CORE module of SQD with MTPS seed. S-CORE(MTPS) differs from SQD
only in terms of the initialization of the configuration recovery module: SQD uses quantum hardware samples for dimeric systems
while S-CORE(MTPS) uses MTPS obtained from monomeric PIGen-SQD solutions.
% In this case SQD refers to the conventional SQD with carryover mechanism for dimeric system
% with dimeric hardware samples. 
% The hyperparameters (such as number of batches, samples drawn per iteration etc.)
% used for SQD and S-CORE (MTPS) iterations to obtain the results are
% given in \textcolor{blue}{SUPPL}.
In this case, one can clearly see the importance of
initial seed for S-CORE module. While SQD
suffers from slow convergence, inaccurate energy estimation and huge diagonalization dimension,
the corresponding S-CORE(MTPS) seed shows faster convergence, provides significantly more accurate results while using
substantially low dimensional diagonalization space for both $\mathrm{(H_2O)_2}$ and FMA dimer (Fig. \ref{fig: coupling map and S-CORE with MTPS} (b) and (d)).
This comparison shows that MTPS, which is obtained from shallower monomeric circuits,
low shot budget and consequently better span of the monomeric Hilbert space, shows
substantial improvements in terms of both accuracy and efficiency compared to standard SQD
even with the same S-CORE mechanism.

The corresponding PIGen-SQD(MAPS) in both the cases provides the perfect balance of accuracy and efficiency.
% in terms of diagonalization subspace dimension.
The systematic iterative identification of dominant configurations and blacklisting mechanism of PIGen-SQD(MAPS)
provides 95\% reduction in diagonalization space dimension while having an order-of-magnitude better energy accuracy for $\mathrm{(H_2O)}_2$.
Somewhat similar trend is observed in $\mathrm{(FMA)_2}$ as well.
% restricts the diagonalization space at each iteration in this study as demonstrated in Fig. \ref{fig: coupling map and S-CORE with MTPS}.

% The corresponding MTPS-PIGen-SQD shows the perfect balance between accuracy and
% compactness of the associated subspace, establishing its efficiency over carryover mechanism.

\textbf{Potential Energy Surface (PES):}
Fig. \ref{fig: pes plot H2O dimer}(a) shows the PES for water dimer obtained with SQD, PIGen-SQD(MAPS) and S-CORE(MTPS)
with varying oxygen-oxygen distance (in \AA) of two different monomers. For comparison, corresponding HCI,
SCI-PySCF and CCSD results are also plotted. Fig. \ref{fig: pes plot H2O dimer}(b) and (c) shows the corresponding error
with respect to HCI and diagonalization dimension respectively. The results show that PIGen-SQD(MAPS),
SCI-PySCF and S-CORE(MTPS) are all within 1 $\mathrm{mE_h}$ accuracy with HCI for all the intermolecular Oxygen-Oxygen distances under consideration.
CCSD has around 2 $\mathrm{mE_h}$ error while SQD shows around 8-10 $\mathrm{mE_h}$ error throughout all the distances.
The interesting observation is that even though SQD and S-CORE(MTPS) uses the same configuration recovery module,
the MTPS initialization steers S-CORE toward exploring the relevant sectors of the Hilbert space and shows remarkable accuracy
while restricting the diagonalization to substantially compact subspaces in comparison with SQD. However, SQD accuracy can be further increased
by tuning the associated hyperparameters albeit with increased diagonalization dimension.
For all the distances, PIGen-SQD(MAPS) provides the most compact diagonalization subspace in this study while attaining
consistently around 0.4 $\mathrm{mE_h}$ accuracy with HCI. Fig. \ref{fig: pes plot formamide dimer}(a) shows the potential energy curve of the FMA dimer as a function of the perpendicular distance between the parallel C--N bonds of the two monomers.
Fig. \ref{fig: pes plot formamide dimer}(b) and (c) demonstrates S-CORE(MTPS) is the most accurate, providing an order-of-magnitude
more accurate energetics compared to conventional SQD while utilizing only one third of the diagonalization cost.
In this case also, PIGen-SQD(MAPS) shows the perfect balance between
accuracy and compact diagonalization space as it consistently has 0.4 $\mathrm{mE_h}$ error over the surface and requires almost
an order-of-magnitude reduced number of configurations in the diagonalization space.
SCI-PySCF lies somewhere in between in terms of accuracy and efficiency.

\begin{table*}[t]
\caption{\textbf{Comparison between RHF, HCI, PIGen-SQD(MAPS) and CCSD under the FMO-2 approximation.}}
\label{tab:fmo2 results}
\centering
% Adjust this length to change the spacing between columns (default is usually 6pt)
\setlength{\tabcolsep}{10pt} 
\begin{tabular*}{\textwidth}{@{\extracolsep{\fill}}lcccc}
\toprule
$E_{\mathrm{RHF}}^{\mathrm{FMO-2}}$ &
$E_{\mathrm{CCSD}}^{\mathrm{FMO-2}}$ & 
$E_{\mathrm{PIGen-SQD(MAPS)}}^{\mathrm{FMO-2}}$ &
$E_{\mathrm{HCI}}^{\mathrm{FMO-2}}$  &
 \\
\midrule
-456.00503267  &  -456.54217401  & -456.54423085   & -456.54514579  \\
\bottomrule
\end{tabular*}
\end{table*}

% \subsection{Water Hexamer}

% To test the distributed workflow for extended systems, we have considered water hexamer $\mathrm{(H_2O)_6}$ with a cage
% structure under the dimeric ($n=2$) or FMO-2 approximation. Under this fragmentation scheme, the system is partitioned into
% $N = 6$ monomers and $\binom{6}{2}=15$ fragment pairs (or dimers).
% All the dimers are treated with the PIGen-SQD(MAPS) algorithm in parallel
% and \enquote{stitched} together with the many-body expansion according to Eq. \eqref{eq: many body expansion}.

\subsection{Water Hexamer}

To test the scalability and accuracy of our distributed workflow for extended systems, we investigated the cage conformer of the water hexamer,
$\mathrm{(H_2O)_6}$, under the two-body fragment molecular orbital (FMO-2) approximation. Within this framework, the total system is partitioned into $N = 6$ monomeric units, yielding $\binom{6}{2} = 15$ unique dimeric pairs (fragment pairs). Each dimer is treated independently in parallel using the \mbox{PIGen-SQD(MAPS)} algorithm, and the individual subsystem energetics are aggregated via the many-body expansion according to Eq.~\eqref{eq: many body expansion QSCI}.

The resulting FMO-2 energies obtained across restricted Hartree--Fock (RHF), heat-bath configuration interaction (HCI), \mbox{PIGen-SQD(MAPS)}, and coupled-cluster with singles and doubles (CCSD) are summarized in Table~\ref{tab:fmo2 results}. Across the 120-qubit active space spanned by the hexamer, direct full-system state preparation and measurement are entirely bypassed; instead, quantum sampling is restricted to 20-qubit monomeric circuits, followed by inter-fragment correlation recovery across 40-qubit dimeric subsystems.

Remarkably, \mbox{PIGen-SQD(MAPS)} achieves an FMO-2 ground-state energy of $-456.544231~E_{\mathrm{h}}$, recovering the
electronic correlation within sub-millihartree accuracy ($0.915~mE_{\mathrm{h}}$) relative to the classical reference $E_{\mathrm{HCI}}^{\mathrm{FMO-2}} = -456.545146~E_{\mathrm{h}}$. Furthermore, \mbox{PIGen-SQD(MAPS)} outperforms $E_{\mathrm{CCSD}}^{\mathrm{FMO-2}} = -456.542174~E_{\mathrm{h}}$ by nearly $2.06~mE_{\mathrm{h}}$, demonstrating the superior capability of the physics-informed generative recovery scheme to capture higher-order inter-fragment correlation effects beyond standard single- and double-excitation manifolds. These results validate \mbox{PIGen-SQD(MAPS)} as an accurate and resource-efficient solver for utility-scale chemical simulations on near-term quantum-HPC architectures.

\section{Conclusion and Future Outlook}
\label{sec:conclusion}

In this work, we propose a distributed quantum-HPC hybrid architecture to treat inter-molecular electronic correlations, larger active spaces and extended systems.
The developed algorithm involves distributed QCSC solvers and multi-layered Gen-ML modules to decompose a large-scale problem into parallel tractable computational tasks,
followed by a two-stage aggregation process to integrate the individual sub-systems and reconstruct the solution of the super-system.
Our algorithm, which we refer to as PIGen-SQD(MAPS), employs PIGen-SQD as the distributed QCSC solver which provides a set of dominant configurations and the associated coefficients
for each sub-system. These individual solutions are collected and fed into a two-stage aggregation process to get the target solution at a much reduced
computational cost compared to the parent QCSC method. The aggregation step starts with a computational bit assignment, which provides an one-to-one mapping between the
local and global bits. Subsequently, a mapped tensor product state (MTPS) is constructed to recover a dominant subspace of the global Hilbert space from the dominant configurations of the individual fragments.
Finally MTPS is fed into an iterative inter-fragment correlation recovery module
driven by Gen-ML or carryover S-CORE mechanism. The multi-layered Gen-ML engine, one at the parallel QCSC level and one at inter-fragment recovery stage
drives the algorithm along with quantum hardware samples (obtained for each fragmented systems) and physically motivated MTPS construction.
Due to the distributed nature, PIGen-SQD(MAPS) requires fewer qubits, shallower circuits, and drastically reduced shot requirements per QCSC task that
improves the quality of the samples obtained for each sub-problem compared to conventional SQD.

Our numerical demonstrations on non-covalent dimeric systems, including the hydrogen-bonded water dimer $\mathrm{(H_2O)_2}$ and the peptide-bond
prototype formamide dimer $(\mathrm{FMA})_2$, reveal that \mbox{PIGen-SQD(MAPS)} yields an order-of-magnitude improvement in energy accuracy compared to standard SQD performed
on the entire composite system, while simultaneously reducing the classical Hamiltonian diagonalization dimension by up to $95\%$. Furthermore, the application to the cage
conformer of the water hexamer $\mathrm{(H_2O)_6}$ under the
FMO-2 approximation illustrates the scalability of our approach, effectively decomposing a demanding 120-qubit problem into six parallel 20-qubit monomeric QCSC tasks, followed
by aggregation stages.
The resulting aggregated subspace shows strong agreement with the reference calculations with HCI and provides better accuracy than reference FMO-2-CCSD calculations.

By restricting expensive quantum measurements to localized active spaces and Gen-ML driven cross-fragment entanglement,
\mbox{PIGen-SQD(MAPS)} provides a more scalable and systematically improvable avenue for electronic structure calculations on utility-scale quantum-HPC architectures. Furthermore, the framework provides us with great flexibility in selectively capturing the many-body correlation among the weakly entangled fragments through MTPS with further cross-fragment entanglement build-up by Gen-ML or S-CORE(MTPS). This hybrid mechanism allows us to treat different 
parts of a large system with various degrees of sophistication as warranted by its electronic complexity. Future extensions will focus on generalizing the bit-mapping protocol to heterogeneous environments and integrating advanced generative architectures such as transformers\cite{zeng2026transformer} or autoencoders\cite{halder2026quantum} for multi-fragment correlation recovery.

% This way, we have addressed the primary computational bottlenecks of SQD namely high-depth quantum circuits, excessive measurement shot budgets, and prohibitive classical diagonalization dimensions - by introducing a distributed quantum-centric supercomputing framework termed \mbox{PIGen-SQD(MAPS)}. By decomposing extended molecular active spaces into chemically intuitive sub-problems, the framework avoids full-system state preparation and restricts direct quantum hardware sampling exclusively to compact, low-qubit fragment subsystems.
% This is followed by an aggregation stage to integrate the individual solutions.  

% The framework achieves its high efficiency through a synergistic two-stage classical post-processing workflow. First, local-to-global orbital correspondence is formulated as a linear assignment problem (LAP), mapping local computational bitstrings onto the global Fock space to construct a mapped tensor product state (MTPS) that retains dominant intra-fragment electronic correlations. Building upon this initialization, a secondary Gen-ML layer captures non-local inter-fragment electron interactions to systematically reconstruct the dominant configuration space of the composite super-system.

\section*{Author contributions}
% We strongly encourage authors to include author contributions and recommend using \href{https://casrai.org/credit/}{CRediT} for standardised contribution descriptions. Please refer to our general \href{https://www.rsc.org/journals-books-databases/journal-authors-reviewers/author-responsibilities/}{author guidelines} for more information about authorship.

% Contributor Roles:
\textbf{Conceptualization}: Chayan Patra, Prayas Ghosh, Rahul Maitra

\textbf{Methodology}: Chayan Patra, Prayas Ghosh, Rahul Maitra

\textbf{Investigation:} Chayan Patra, Prayas Ghosh, Rahul Maitra

\textbf{Formal analysis:} Chayan Patra, Prayas Ghosh, Rahul Maitra

\textbf{Data curation:} Chayan Patra, Prayas Ghosh

\textbf{Writing - original draft:} Chayan Patra, Rahul Maitra

\textbf{Writing - review and editing:} Chayan Patra, Prayas Ghosh, Rahul Maitra

\textbf{Software:} Chayan Patra, Prayas Ghosh, Adwait Patkhedkar

\textbf{Visualization:} Chayan Patra, Prayas Ghosh

\textbf{Supervision:} Rahul Maitra

\textbf{Funding acquisition:} Rahul Maitra

% C. P. and R. M. led the conceptualization, methodology, investigation and formal analysis of the study. C. P. carried out the data curation and code development as well as drafted the original manuscript. R. M. supervised the research, secured funding, and co-wrote and revised the manuscript. D. M. supported the investigation and formal analysis. M. R. L., S. H., R. G. and D. H. contributed to software development and assisted with manuscript revision.

% \section*{AUTHOR DECLARATIONS}
% \section*{Competing Interests}
\section*{CONFLICT OF INTEREST}
The authors declare no conflict of interest.

\section*{Data and Code Availability}
The data and code are available upon reasonable request to the corresponding author.

\section*{Acknowledgments}

We acknowledge the use of IBM Quantum Credits for this work. The
views expressed are those of the authors, and do not reflect the official
policy or position of IBM or the IBM Quantum team.
% The authors acknowledge Ritajit
% Majumdar from the Quantum Algorithm Engineering team at IBM for fruitful discussions, and contributions to achieving results on quantum hardware.
RM and CP acknowledges the financial support from
Quantum Algorithms project with sanction number DST/QTC/NQM/QA/2026/01
under the National Quantum Mission (NQM) scheme,
Department of Science and Technology (DST), Government of India.
AP and RM acknowledge financial support from Industrial Research and Consultancy Centre (IRCC), IIT Bombay.
The authors acknowledge the discussions with Sonaldeep Halder
on some technical aspects of the code during the initial phase of the project.

\appendix

\counterwithin{figure}{section}

\renewcommand{\thefigure}{\thesection\arabic{figure}}

%\section*{Appendix}

\section{Restricted Boltzmann Machines for Configuration Recovery} \label{appendix: RBM}

Restricted Boltzmann Machines (RBMs) constitute a class of unsupervised, generative stochastic neural networks capable of modeling complex, high-dimensional nonlinear probability distributions. Structurally, an RBM is organized as a bipartite graph comprising a visible layer $\boldsymbol{g} = (\{g_i\}_{i=1}^D, g_i \in \{0,1\})$ that encodes system configurations, and a hidden layer $\boldsymbol{h} = (\{h_j\}_{j=1}^J, h_j \in \{0,1\})$ that introduces auxiliary degrees of freedom to enhance expressivity. The network parameters $\theta = \{\mathbf{u}, \mathbf{y}, \mathbf{W}\}$
consisting of the visible bias vector $\mathbf{u}$, the hidden bias vector $\mathbf{y}$, and the inter-layer weight matrix $\mathbf{W}$, are optimized to accurately approximate the underlying data distribution. 

Mathematically, the joint probability distribution over the visible and hidden units is governed by an energy function:
\begin{equation}
    E(\boldsymbol{g}, \boldsymbol{h}, \theta) = -\sum_{i=1}^D \sum_{j=1}^J g_i W_{ij} h_j - \sum_{j=1}^J y_j h_j - \sum_{i=1}^D u_i g_i,
\end{equation}
from which the marginal probability of a visible configuration $\boldsymbol{g}$ is derived via the canonical Gibbs distribution:
\begin{equation}
    P(\boldsymbol{g}; \theta) = \sum_{\boldsymbol{h}} P(\boldsymbol{g}, \boldsymbol{h}; \theta) = \sum_{\boldsymbol{h}} \frac{1}{Z(\theta)} \exp\left(-E(\boldsymbol{g}, \boldsymbol{h}, \theta)\right),
\end{equation}
where the partition function $Z(\theta)$ normalizes the distribution over all possible states:
\begin{equation}
    Z(\theta) = \sum_{\boldsymbol{g}} \sum_{\boldsymbol{h}} \exp\left(-E(\boldsymbol{g}, \boldsymbol{h}, \theta)\right).
\end{equation}

\begin{table}[t]
\caption{\label{tab:thesholds_and_hyperparameters}Summary of thresholds and hyperparameters used in PIGen-SQD workflow.}
\begin{ruledtabular}
\begin{tabular}{ll}
Thresholds and Hyperparameters & Value \\
\colrule
Two-Body Integral Threshold ($\epsilon_{int}$) & $10^{-10}$ \\
% RBM generation &
% $\mathcal{O}(N_{\mathrm{gen}}DJ)$ \\
Blacklist Screening Coeff.\ ($\epsilon_{\mathrm{coeff}}$) & $10^{-6}$ \\
PIGen-SQD Convergence Threshold ($\epsilon_E$)  & $10^{-5}$ \\
RBM Learning Rate ($\eta$) & $10^{-3}$ \\
RBM Intrinsic Gibbs Sampling ($n_G$) & 20 \\
RBM Number of Training Epochs ($n_{\mathrm{train}}$) & 3 \\
RBM Generation Gibbs Sampling ($n_g$) & 1 \\
\end{tabular}
\end{ruledtabular}
\end{table}

Because evaluating the partition function exactly is computationally intractable, RBM training typically relies on the contrastive divergence (CD) algorithm\cite{hinton2012practical}. CD performs stochastic approximations of the log-likelihood gradient, effectively minimizing the Kullback--Leibler divergence between the empirical data distribution and the model's learned distribution. Owing to the absence of intra-layer connections, the conditional activation probabilities update alternately via the logistic sigmoid function $\sigma(x) = (1 + e^{-x})^{-1}$:
\begin{align}
    P(h_j = 1 \mid \boldsymbol{g}) &= \sigma\left(\sum_{i=1}^D g_i W_{ij} + y_j\right), \\
    P(g_i = 1 \mid \boldsymbol{h}) &= \sigma\left(\sum_{j=1}^J W_{ij} h_j + u_i\right).
\end{align}
Following training, novel samples are drawn according to the learned probability distribution $P(\boldsymbol{g})$ using the block Gibbs sampling technique\cite{mackay2003information}.
Detailed descriptions of RBM can be found in Refs.\cite{hinton2012practical, zhang2018overview, hinton2002training, hinton2006reducing, melko2019restricted}.

In the domain of electronic structure theory, RBMs offer exceptional representational capacity for capturing intricate many-body electron correlations\cite{herzog2023solving, xia2018quantum, singh2025maximal, halder2024machine, halder2025construction, hernandez2025configuration}. When properly trained, an RBM directs configuration sampling toward physically significant sectors of the Hilbert space consisting of determinants possessing the largest configuration interaction (CI) amplitudes. Consequently, the diagonalization subspace drastically shrinks and computational
overhead is mitigated. Nevertheless, as unsupervised architectures, RBMs are inherently sensitive to the quality of the initial visible layer training data.
In this paper, for PIGen-SQD, we have initialized RBM training with hardware samples augmented by CISD configurations and perturbatively obtained triples (see Appendix \ref{appendix: PT algebraic terms}). The second layer of RBM for PIGen-SQD(MAPS) workflow uses MTPS as the initial training data set. The details of the associated hyperparameters used in this work is given in
Table \ref{tab:thesholds_and_hyperparameters} which is same as the original PIGen-SQD work in Ref. \cite{Patra_2026}.

\section{Hyperparameters for SQD and S-CORE(MTPS)} \label{app: hyperparameters for SQD and S-CORE(MTPS)}

The hyperparameters for iterative carryover mechanism of SQD and S-CORE(MTPS) are given in Table \ref{tab:SQD_hyperparameters}.
The two methods use the same set of hyperparameters, except for the number of samples per batch.

\begin{table}[t]
\caption{\label{tab:SQD_hyperparameters} Hyperparameters used in SQD and S-CORE(MTPS).}
\begin{ruledtabular}
\begin{tabular}{lll}
Hyperparameters & SQD & S-CORE(MTPS) \\
\colrule
Iterations & 10 & 10 \\
Number of Batches & 2 & 2 \\
Samples Per Batch & 1000 & 10000 \\
Energy Tolerance & $10^{-5}$ & $10^{-5}$ \\
Occupancies Tolerance & $10^{-5}$ & $10^{-5}$ \\
Carryover Threshold & $10^{-6}$ & $10^{-6}$ \\
% RBM generation &
% % $\mathcal{O}(N_{\mathrm{gen}}DJ)$ \\
% Blacklist Screening Coeff.\ ($\epsilon_{\mathrm{coeff}}$) & $10^{-6}$ \\
% RBM Recovery Convergence Threshold ($\epsilon_E$)  & $10^{-5}$ \\
% RBM Learning Rate ($\eta$) & $10^{-3}$ \\
% RBM Intrinsic Gibbs Sampling ($n_G$) & 20 \\
% RBM Number of Training Epochs ($n_{\mathrm{train}}$) & 3 \\
% RBM Generation Gibbs Sampling ($n_g$) & 1 \\
\end{tabular}
\end{ruledtabular}
\end{table}

\section{THE LINEAR ASSIGNMENT PROBLEM} \label{appendix sec: LAP}

To determine the optimal pairing between two distinct sets of equal size, such as mapping initial states to final states, assigning tasks to workers of a company for optimal output or assigning discrete particles to lattice sites, we formulate the system as a Linear Assignment Problem (LAP). The LAP is resolved exactly and deterministically in polynomial time using the Kuhn-Munkres optimization procedure, commonly known as the Hungarian algorithm.

\subsection{Mathematical Formulation of LAP}

Consider a bipartite system comprising two disjoint sets of size $n$, denoted $\mathcal{U}$ and $\mathcal{V}$. Let $C \in \mathbb{R}^{n \times n}$ denote the cost matrix, where the matrix element $c_{ij}$ represents the physical cost, energy, or distance associated with assigning element $i \in \mathcal{U}$ to element $j \in \mathcal{V}$. 

The objective is to find a bijective mapping that minimizes the global cost function. We introduce a boolean decision matrix $X \in \{0,1\}^{n \times n}$, where $x_{ij} = 1$ if $i$ is assigned to $j$, and $x_{ij} = 0$ otherwise. The optimization problem is mathematically defined as:

\begin{equation}
\min_{X} \sum_{i=1}^{n} \sum_{j=1}^{n} c_{ij} x_{ij}
\end{equation}
subject to the constraints of a perfect bipartite matching (each element in $\mathcal{U}$ is mapped to exactly one element in $\mathcal{V}$, and vice versa):

\begin{align}
\sum_{j=1}^{n} x_{ij} &= 1 \quad \forall i \in \{1, 2, \dots, n\} \\
\sum_{i=1}^{n} x_{ij} &= 1 \quad \forall j \in \{1, 2, \dots, n\}
\end{align}

\subsection{Algorithmic Implementation of Hungarian Method}

The Hungarian algorithm exploits the property that adding or subtracting a constant from any row or column of the cost matrix shifts the total objective function by a constant, but preserves the configuration of the optimal assignment. The algorithm systematically reduces the matrix $C$ to induce a zero-cost perfect matching. The procedure operates in $\mathcal{O}(n^3)$ time and executes as follows:

\paragraph{Matrix Reduction}
First, we normalize the cost matrix by establishing local minima in every row and column. For each row $i$, we subtract the minimum element from all entries in that row:
\begin{equation}
c'_{ij} = c_{ij} - \min_{k} c_{ik} \quad \forall j
\end{equation}
Subsequently, for each column $j$ in the intermediate matrix $C'$, we subtract the column minimum:
\begin{equation}
c''_{ij} = c'_{ij} - \min_{k} c'_{kj} \quad \forall i
\end{equation}
The resulting reduced matrix $C''$ contains at least one zero in every row and column.

\paragraph{Graph Construction and Vertex Cover}
We define an unweighted bipartite graph $\mathcal{G} = (\mathcal{U}, \mathcal{V}, \mathcal{E})$ where an edge $(i, j) \in \mathcal{E}$ exists if and only if $c''_{ij} = 0$. By K\"onig's theorem, the size of the maximum bipartite matching in $\mathcal{G}$ is exactly equal to the minimum vertex cover. 

We determine the minimum number of lines---corresponding to a set of rows $\mathcal{R}$ and a set of columns $\mathcal{C}$---required to cover all zero elements in $C''$. Let $k = |\mathcal{R}| + |\mathcal{C}|$ be the total number of covering lines.

\paragraph{Optimality Test}
\begin{itemize}
    \item If $k = n$, the zero elements covered by these lines constitute a perfect matching. The algorithm terminates, and the optimal assignment matrix $X$ is constructed from a set of $n$ independent zeros.
    \item If $k < n$, a perfect matching cannot yet be formed using purely zero-cost edges. The matrix must be iteratively updated to introduce new zeros.
\end{itemize}

\paragraph{Matrix Update}
To iteratively approach optimality without discarding previously discovered partial matchings, we define $\Delta$ as the minimum element in $C''$ that is not covered by any line in $\mathcal{R}$ or $\mathcal{C}$:
\begin{equation}
\Delta = \min_{i \notin \mathcal{R}, \, j \notin \mathcal{C}} c''_{ij}
\end{equation}

The matrix elements are updated according to the following piecewise transformation:
\begin{equation}
c^{(new)}_{ij} = 
\begin{cases} 
c''_{ij} - \Delta, & \text{if } i \notin \mathcal{R} \text{ and } j \notin \mathcal{C} \text{ (uncovered)} \\ 
c''_{ij} + \Delta, & \text{if } i \in \mathcal{R} \text{ and } j \in \mathcal{C} \text{ (doubly covered)} \\ 
c''_{ij}, & \text{otherwise (singly covered)} 
\end{cases} 
\end{equation}

We substitute $C''$ with $C^{(new)}$ and repeat the procedure from Step 2 until condition $k = n$ is satisfied.

\section{Fragment-to-Supersystem Computational Bit Mapping and Cost Function Construction} \label{appendix sec: Orbital Mapping and Cost Function}
The construction of MTPS from configurations obtained from fragment QCSC solvers requires a consistent correspondence between the fragment molecular orbitals and the supersystem molecular orbitals. The correspondence is non-trivial because the supersystem orbitals are not required to retain the same ordering or strict fragment localization as the isolated-fragment orbitals. Orbital mixing, energetic reordering and near degeneracies may therefore render a direct index-by-index correspondence. To obtain a one to one correspondence, the fragment to supersystem mapping is constructed as linear assignment problem (LAP). LAP with Hungarian method
% and Jonker-Volgenant solution process
is described in the Appendix ~\ref{appendix sec: LAP}. In this section, we are describing the construction of physically motivated cost function supplied to the assignment solver for the Fragment-to-Supersystem Orbital Mapping.

\subsection{Orbital Overlap Similarity}
The spatial correspondence between supersystem MO $\tilde{\chi}_g$ and fragment MO $\chi_{\beta_k}$ is 
\begin{equation}
    S_{g\beta_k}= \langle \tilde{\chi}_g \mid \chi_{\beta_k} \rangle
\end{equation}
% where both $\chi_i$ and $\chi_{I_\beta}$ can be written as linear combination of atomic orbitals. 
% \begin{equation} \label{AO to MO eqn}
%     \chi_{i} = \sum_{\nu} \tilde{c}_{\nu i} \xi_{\nu}\hspace{5mm}
%     \chi_{I_\beta} = \sum_{\mu} \tilde{c}_{\mu I_\beta} \xi_{\mu}^I,  
% \end{equation}
% Here important to note that as we have embedded the fragment orbital coefficients in the supersystem AO-basis we can use same AO-basis to construct both fragment and supersystem Molecular orbitals. For a fragment molecular orbital $\chi_{I_\beta}$, $\tilde{c}_{\mu I_{\beta}}$ will be zero for those supersystem atomic orbitals that do not correspond to that particular fragment system. So we can rewrite equation \ref{AO to MO eqn} as
% \begin{equation} 
%     \chi_{i} = \sum_{\nu}^{N_{AO}^{sup}} \tilde{c}_{\nu i} \xi_{\nu}\hspace{5mm}
%     \chi_{I_\beta} = \sum_{\mu}^{N_{AO}^{sup}} \tilde{c}_{\mu I_{\beta}} \xi_{\mu},  
% \end{equation}
Plugging in the expansion in Eq. \eqref{eq: Ao to MO expansion} we have
\begin{equation}
    S_{g\beta_k}=\sum_{\nu} \sum_{\mu} \tilde{C}_{g \nu} \langle \tilde{\xi}_{\nu} | \xi_{\mu_k} \rangle {C}_{\mu_k \beta_k} 
\end{equation}
Here, it is important to note that computationally there is a mismatch between the dimension of $\tilde{C}_{g \nu}$ and $C_{ \mu \beta_k}$,
which is taken care of by padding extra zero coefficients before evaluating the overlap for each block involving a particular $k$ of the entire $S$ matrix.

As the global phase of a molecular orbital is arbitrary and consequently any two orbitals differing only by a global phase have identical mapping similarity, we define
the spatial similarity factor as the square of the orbital similarity,
\begin{equation} \label{Overlap_compatibility}
    A_{g{\beta_k}} =  |S_{g\beta_k}|^2
\end{equation}
For normalized orbitals, $0\leq A_{g{\beta_k}} \leq 1$ where values approaching unity indicate strong spatial overlap, whereas values close to zero reflect negligible spatial correspondence between the fragment and supersystem molecular orbitals.

\subsection{Orbital Energy Compatibility}
Spatial similarity alone may not always uniquely distinguish the physically appropriate correspondence. This particularly becomes more relevant for groups of closely spaced orbitals having similar spatial character or for orbitals that mix upon formation of the supersystem. Therefore, we introduced orbital energy approximately as an additional compatibility criterion.

For $g$-th global orbital with orbital energy $\epsilon_g$ and $\beta$-th orbital of the $k$-th fragment is $\epsilon_{\beta_k}$ we defined energy compatibility factor as a Gaussian function
% \begin{equation}
%     (\Delta \epsilon)_{g \beta_k} = e ^{ - \frac{(\epsilon_g - \epsilon_{\beta_k }) ^2}{2 \sigma ^2}} 
% \end{equation}
\begin{equation}
    (\Delta \epsilon)_{g \beta_k} = e ^{ - \frac{(\epsilon_g - \epsilon_{\beta_k} - \overline{\delta\epsilon}_{\beta_k}) ^2}{2 \sigma_E ^2}} 
\end{equation}
here $\sigma$ controls the energy window over which fragment and supersystem orbitals are considered energetically compatible. The Gaussian factor satisfies $0 \leq (\Delta \epsilon)_{g \beta_k} \leq 1$ and smoothly penalizes increasingly large energy separations without introducing a hard energy cutoff.  

\subsection{System dependent calibrated energy window}
The energy compatibility factor introduced above needs an energy scale $\sigma$ that determines how fast the compatibility between a fragment orbital and a supersystem orbital decreases with increase in orbital energy separation. A fixed value of $\sigma$ would mean same absolute energy tolerance on chemically and spectrally different systems and thus introduce a system independent scale into the mapping. Instead, we determined $\sigma$ directly through the overlap-resolved supersystem spectrum associated with the fragment orbitals.
First we made the overlap distribution normalized according to 
\begin{equation}
p_{g\beta_k}
=
\frac{A_{g\beta_k}}
{\displaystyle\sum_f A_{f\beta_k}},
\qquad
\sum_g p_{g\beta_k}=1
\label{eq:normalized_overlap_weight}
\end{equation} where $A_{g\beta_k}$ is defined as equation 
\ref{Overlap_compatibility}. The weights $p_{g\beta_k}$ is therefore identify the regions of the supersystem spectrum that actually carry the character of the fragment orbital $\chi_{\beta_k}$. Now we have the overlap-conditioned mean energy separation as 
\begin{equation}
\overline{\delta\epsilon}_{\beta_k}
=
\sum_g
p_{g\beta_k}
\delta\epsilon_{g\beta_k}
\label{eq:overlap_conditioned_shift}
\end{equation}
where $\delta\epsilon_{g\beta_k} = \epsilon_g - \epsilon_{\beta_k}$,
and the corresponding spectral spread as 
\begin{equation}
\tau_{\beta_k}
=
\left[
\sum_g
p_{g\beta_k}
\left(
\delta\epsilon_{g\beta_k}
-
\overline{\delta\epsilon}_{\beta_k}
\right)^2
\right]^{1/2}
\label{eq:overlap_conditioned_spread}
\end{equation}
The quantity in equation \ref{eq:overlap_conditioned_shift} measures a systematic displacement of the fragment orbital relative to the supersystem spectrum (i.e. it finds out location of the overlap-supported supersystem orbitals carrying the character of the fragment orbital), while the quantity of the equation \ref{eq:overlap_conditioned_spread} describes the energy range over which orbital characters are spread (the intrinsic width of this overlap-supported distribution).

% For each fragment orbital $I_\beta$, combination of these two quantity give the overlap-conditioned spectral scale
% \begin{equation}
% s_{I_\beta}
% =
% \left[
% \left(
% \overline{\Delta\epsilon}_{I_\beta}
% \right)^2
% +
% \tau_{I_\beta}^2
% \right]^{1/2}
% \label{eq:fragment_spectral_scale}
% \end{equation}
% Using the definitions in equation \ref{eq:overlap_conditioned_shift} and \ref{eq:overlap_conditioned_spread} we get
% \begin{equation}
% s_{I_\beta}
% =
% \left[
% \sum_i
% p_{iI_\beta}
% \left(
% \Delta\epsilon_{iI_\beta}
% \right)^2
% \right]^{1/2}
% \label{eq:fragment_spectral_second_moment}
% \end{equation}
% Therefore, $s_{I_\beta}$ is the overlap-conditioned root-mean-square energy displacement of fragment orbital $I_\beta$ within the supersystem orbital manifold.

The single characteristic energy scale for the overall fragment to supersystem mapping problem is obtained from the root mean square of the orbital specific scales,
\begin{equation}
\tau_{\mathrm{spec}}
=
\left[
\frac{1}{N_{\mathrm{frag}}}
\sum_{\beta_k}
\tau_{\beta_k}^{\,2}
\right]^{1/2},
\label{eq:spectral_energy_scale}
\end{equation}
where $N_{\mathrm{frag}}$ is the total number of active fragment orbitals included in the mapping.
The width of the gaussian energy compatibility factor is then defined by 
% \begin{equation}
%     \sigma_E = \kappa s_{spec}
% \end{equation}
\begin{equation}
    \sigma_E =  \tau_{spec}
\end{equation}
% where $\kappa$ is a dimensionless broadening parameter. 
and the energy compatibility factor becomes full defined as
\begin{equation}
    (\Delta \epsilon)_{g \beta_k} = e ^{ - \frac{(\epsilon_g - \epsilon_{\beta_k} - \overline{\delta\epsilon}_{\beta_k}) ^2}{2 \sigma_E ^2}} 
\end{equation}

\subsection{Combined Orbital Mapping Cost Function}
For the overall cost function we combined the spatial similarity factor and energy penalty factor multiplicatively such that the joint compatibility factor is
\begin{equation}
        Q_{g{\beta_k}} = A_{g{\beta_k}} (\Delta \epsilon)_{g \beta_k}
\end{equation}
and the cost function becomes 
\begin{equation}
    J_{g{\beta_k}} = 1- Q_{g{\beta_k}} %[A_{g{\beta_k}} (\Delta \epsilon)_{g \beta_k}]
\end{equation}
% or explicitly,
% \begin{equation}
% J_{g{\beta_k}}
% =
% 1-\left|\sum_{\nu} \sum_{\mu} \tilde{c}_{\nu i} \langle\xi_{\nu} \mid  \xi_{\mu} \rangle \tilde{c}_{\mu I_{\beta}}\right|^2
% e ^{ - \frac{(\epsilon_i - \epsilon_{I_\beta }) ^2}{2 \sigma_E ^2}} 
% \label{eq:full_mapping_cost}
% \end{equation}
The multiplicative construction is deliberately different from an additive combination of  independently weighted spatial and energetic penalties. The two factors here are not competing objectives (whose relative weight are to be adjusted through empirical coefficients), rather they provide complementary conditions for identifying the same physical correspondence i.e. the orbitals should have compatible spatial character and the energy of the spatially related states should be spectrally reasonable. 
In this aspect, it is
important to note that, because \begin{equation}
0\leq A_{g\beta_k}\leq1,
\qquad
0<(\Delta \epsilon)_{g \beta_k}\leq1,
\end{equation}
the joint compatibility satisfies 
\begin{equation}
0\leq
Q_{g\beta_k}
\leq
A_{g\beta_k}.
\label{eq:compatibility_bound}
\end{equation}
This ensures that energy proximity can never overrule the absence of the spatial compatibility. It can only reduce the compatibility already supported by the spatial similarity or at most keep it preserved. 
Thus, if
\begin{equation}
A_{g\beta_k}\rightarrow0,
\end{equation}
then
\begin{equation}
Q_{g\beta_k}\rightarrow0
\qquad\text{and}\qquad
J_{g\beta_k}\rightarrow1
\end{equation}
for any value of $(\Delta\epsilon)_{g\beta_k}$. An accidental near degeneracy in orbital energy therefore cannot compensate for negligible spatial similarity.
On the other hand, when two orbitals exhibit strong spatial similarity,
\begin{equation}
A_{g\beta_k}\rightarrow1
\quad\Longrightarrow\quad
J_{g\beta_k}\rightarrow1-(\Delta \epsilon)_{g\beta_k}.
\end{equation}
and here the orbital energy difference serves as additional distinguishing factor between the competing high overlap candidates. 

The cooperative character of the two factors can also be expressed directly through
\begin{equation}
\left.\frac{\partial J_{g\beta_k}}
{\partial A_{g\beta_k}} \right|_{(\Delta \epsilon)_{g\beta_k}}
=
-(\Delta \epsilon)_{g\beta_k}
\leq0,
\end{equation}
and
\begin{equation}
\left.\frac{\partial J_{g\beta_k}}
{\partial (\Delta \epsilon)_{g\beta_k}}\right|_{A_{g\beta_k}}
=
-A_{g\beta_k}
\leq0.
\end{equation}
Thus, an increase in either spatial similarity or energy compatibility can only decrease the cost; improvement in one descriptor never produces a penalty in the other. In this precise sense, overlap and energy similarities act cooperatively rather than antagonistically.

Another very important point to note that, the cost can be written as 

\begin{equation}
J_{g\beta_k}
=
\left(1-A_{g\beta_k}\right)
+
A_{g\beta_k}
\left(1-(\Delta \epsilon)_{g\beta_k}\right),
\label{eq:cost_decomposition}
\end{equation}
which illustrates that the first contribution represents the spatial mismatch, whereas the second contribution represents the energy penalty. But the energy penalty is weighted by the spatial similarity itself. It shows that energetic disagreement is relevant only to the extent that the
candidate pair already shares common orbital character.

\color{black}

\section{Mapped Orbitals for FMA Dimer} \label{appendix: formamide mapped orbitals}

\begin{figure}
    \centering
    \includegraphics[width=1.0\linewidth]{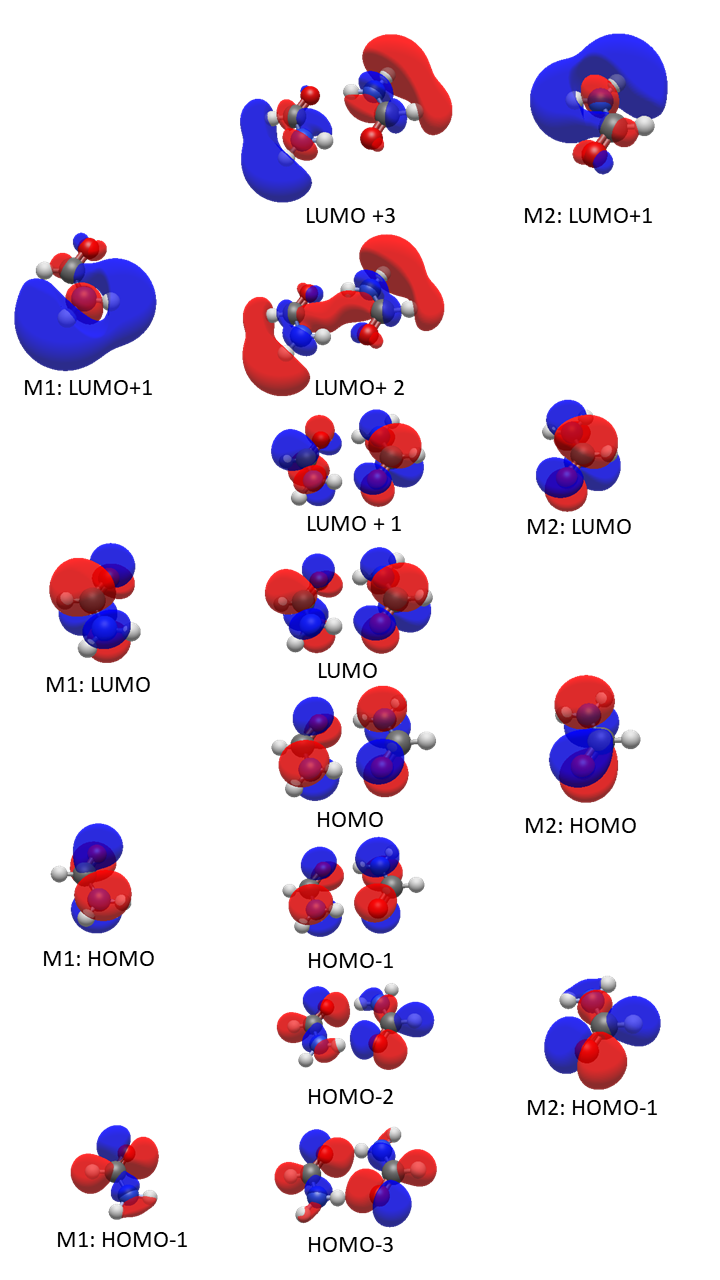}
    \caption{Orbital correspondence in FMA dimer. (Center) Subset of dimer molecular orbitals situated near the HOMO-LUMO region. (Left, Right) Mapped monomeric orbitals for monomer 1 (M1) and monomer 2 (M2), respectively, established through computational bit assignment.
    % Occupied and virtual orbitals of monomer-1 (right) and monomer-2 (left), lying close to the HOMO--LUMO window, are shown together with their corresponding supersystem orbitals (center).(\textcolor{red!65!black}{need horizontal lines to show one to one assignments}. Fragment orbitals (optimized in the field of external potential created by the surrounding fragment.) are embedded in the supersystem atomic-orbital basis, and the assignment cost combines the squared AO-metric orbital overlap with a Gaussian orbital-energy compatibility factor.) The orbitals are shown with same isosurface value and molecular orientations. Red and blue colour denotes the global phase of the orbitals; the overall phase is arbitrary and does not affect the mapping because square of the orbital orbital ($|O_{ij}|^2$) is used in the assignment.
    }
    \label{fig:computational basis mapping FMA dimer}
\end{figure}

Fig. \ref{fig:computational basis mapping FMA dimer} shows a set of super-system orbitals near the HOMO-LUMO region along with the associated assigned fragment orbitals for monomer 1 and monomer 2
according to the computational bit mapping.

\section{Algebraic Terms for Perturbative Calculations} \label{appendix: PT algebraic terms}
The perturbative augmentation in the PIGen-SQD framework configurations up to rank three includes the following algebraic terms:
\begin{equation} \label{tensor contraction terms for triples}
\begin{split}
    & c_{ijk}^{abc} \leftarrow \{v_{ij}^{am}t_{mk}^{bc}, v_{ie}^{ab}t_{jk}^{ec}\}
\end{split}
\end{equation}
% \begin{equation} \label{eq: c4 terms}
% \begin{split}
%     c_{ijkl}^{abcd} \leftarrow & \{v_{ij}^{am} v_{mk}^{bn} t_{nl}^{cd},  v_{ie}^{ab} v_{jf}^{ec} t_{kl}^{fd}, v_{ij}^{am} v_{mf}^{bc} t_{kl}^{fd},  v_{ie}^{ab} v_{jk}^{en} t_{nl}^{cd},\\
%     &  v_{ij}^{am} t_{mn}^{bc} v_{kl}^{nd},  v_{ie}^{ab} t_{jk}^{ef} v_{fl}^{cd},  v_{ij}^{am} t_{mk}^{bf} v_{fl}^{cd}\}
% \end{split}
% \end{equation}
Here $v$ and $t$ represent the two-body integral and MP2 tensor.
Note the signs of the algebraic terms are not explicitly shown in these expressions are assumed to be absorbed within each term.

% \section{Geometry of the Water Hexamer}
% \label{app:water_hexamer}

% The Cartesian coordinates of the cage-shaped water hexamer,
% $\mathrm{(H_2O)_6}$, used in this work are reported in
% Table~\ref{tab:water_hexamer_geometry}. All coordinates are
% given in \AA.

% \begin{table}[htbp]
% \centering
% \caption{Cartesian coordinates of the cage-shaped water hexamer
% $\mathrm{(H_2O)_6}$.}
% \label{tab:water_hexamer_geometry}
% \begin{tabular}{crrr}
% \hline
% Atom & $x$ (\AA) & $y$ (\AA) & $z$ (\AA) \\
% \hline
% O  &  0.67982 &  1.66847 &  0.22469 \\
% H  &  1.54472 &  1.23557 &  0.26190 \\
% H  &  0.80199 &  2.60883 &  0.28673 \\
% O  & -0.92115 & -0.00176 &  1.65020 \\
% H  & -0.44806 &  0.77331 &  1.33150 \\
% H  & -1.76238 & -0.05171 &  1.18135 \\
% O  &  0.63241 & -1.64984 &  0.20904 \\
% H  &  0.08923 & -1.22268 &  0.89198 \\
% H  &  0.53282 & -2.59214 &  0.27953 \\
% O  & -0.42478 &  0.02956 & -1.62376 \\
% H  & -0.09616 &  0.81284 & -1.17004 \\
% H  &  0.00023 & -0.73782 & -1.22438 \\
% O  &  2.74478 & -0.02765 & -0.08537 \\
% H  &  2.20281 & -0.78880 &  0.15326 \\
% H  &  2.94467 & -0.09143 & -1.01363 \\
% O  & -2.76330 & -0.07803 & -0.30968 \\
% H  & -2.08455 & -0.14812 & -0.99436 \\
% H  & -3.30763 &  0.67610 & -0.50478 \\
% \hline
% \end{tabular}
% \end{table}

% \bibliography{apssamp}% Produces the bibliography via BibTeX.

\section*{References}
\providecommand{\noopsort}[1]{}\providecommand{\singleletter}[1]{#1}%
%

% \clearpage
% \onecolumngrid            % if using revtex two-column
% \section*{Supplementary Information}

\clearpage
\onecolumngrid
\begin{center}
  {\large\bfseries Supplementary Information}
\end{center}

% \setcounter{table}{0}\renewcommand{\thetable}{S\arabic{table}}

% reset numbering for the SI
\setcounter{section}{0}
\renewcommand{\thesection}{S\arabic{section}}
\renewcommand{\thesubsection}{\thesection.\arabic{subsection}}
\setcounter{figure}{0}\renewcommand{\thefigure}{S\arabic{figure}}
\setcounter{table}{0}\renewcommand{\thetable}{S\arabic{table}}
\setcounter{equation}{0}\renewcommand{\theequation}{S\arabic{equation}}

\section{Geometry of the Molecular Systems Studied}

The Cartesian coordinates of the water dimer, $\mathrm{(H_2O)_2}$,
with an oxygen--oxygen separation of $
2.2\,\text{\AA}$ are reported in Table~\ref{tab:water_dimer_geometry}.
All coordinates are given in \AA.

\begin{table}[htbp]
\centering
\caption{Cartesian coordinates of the water dimer
$\mathrm{(H_2O)_2}$ with an oxygen-oxygen separation of
$2.2\,\text{\AA}$.}
\label{tab:water_dimer_geometry}
\begin{tabular}{crrr}
\toprule
Atom & $x$ (\AA) & $y$ (\AA) & $z$ (\AA) \\
\midrule
O &  0.63877561 & -0.01416542 &  0.04685218 \\
H &  1.04765561 & -0.58046842 & -0.58807482 \\
H &  0.91879961 &  0.86313358 & -0.16095782 \\
O & -1.55521200 &  0.03613800 & -0.10770500 \\
H & -0.61262100 & -0.04017400 & -0.04964000 \\
H & -1.88509500 & -0.23799800 &  0.73159800 \\
\bottomrule
\end{tabular}
\end{table}

The Cartesian coordinates of the formamide dimer at an intermolecular
separation of $3.67$ {\AA} are reported
in Table~\ref{tab:formamide_dimer_geometry}. All coordinates are
given in \AA.

\begin{table}[htbp]
\centering
\caption{Cartesian coordinates of the formamide dimer at
$3.67$ {\AA}.}
\label{tab:formamide_dimer_geometry}
\begin{tabular}{crrr}
\toprule
Atom & $x$ (\AA) & $y$ (\AA) & $z$ (\AA) \\
\midrule
C & -2.02432754 &  0.05333677 &  0.00000000 \\
N & -1.41581113 & -1.14338767 &  0.00000000 \\
O & -1.45710364 &  1.14749643 &  0.00000000 \\
H & -0.38812535 & -1.20905356 & -0.00000000 \\
H & -1.96870376 & -1.98615669 &  0.00000000 \\
H & -3.12868649 & -0.00951498 &  0.00000000 \\
O &  1.58210808 & -1.14749705 & -0.00000000 \\
C &  2.14932887 & -0.05333586 & -0.00000000 \\
N &  1.54080934 &  1.14338688 &  0.00000000 \\
H &  0.51312326 &  1.20904999 & -0.00000000 \\
H &  3.25368790 &  0.00951875 &  0.00000000 \\
H &  2.09370045 &  1.98615699 &  0.00000000 \\
\bottomrule
\end{tabular}
\end{table}

The Cartesian coordinates of the cage-shaped water hexamer,
$\mathrm{(H_2O)_6}$, used in this work are reported in
Table~\ref{tab:water_hexamer_geometry}. All coordinates are
given in \AA.

\begin{table}[htbp]
\centering
\caption{Cartesian coordinates of the cage-shaped water hexamer
$\mathrm{(H_2O)_6}$.}
\label{tab:water_hexamer_geometry}
\begin{tabular}{crrr}
\toprule
Atom & $x$ (\AA) & $y$ (\AA) & $z$ (\AA) \\
\midrule
O  &  0.67982 &  1.66847 &  0.22469 \\
H  &  1.54472 &  1.23557 &  0.26190 \\
H  &  0.80199 &  2.60883 &  0.28673 \\
O  & -0.92115 & -0.00176 &  1.65020 \\
H  & -0.44806 &  0.77331 &  1.33150 \\
H  & -1.76238 & -0.05171 &  1.18135 \\
O  &  0.63241 & -1.64984 &  0.20904 \\
H  &  0.08923 & -1.22268 &  0.89198 \\
H  &  0.53282 & -2.59214 &  0.27953 \\
O  & -0.42478 &  0.02956 & -1.62376 \\
H  & -0.09616 &  0.81284 & -1.17004 \\
H  &  0.00023 & -0.73782 & -1.22438 \\
O  &  2.74478 & -0.02765 & -0.08537 \\
H  &  2.20281 & -0.78880 &  0.15326 \\
H  &  2.94467 & -0.09143 & -1.01363 \\
O  & -2.76330 & -0.07803 & -0.30968 \\
H  & -2.08455 & -0.14812 & -0.99436 \\
H  & -3.30763 &  0.67610 & -0.50478 \\
\bottomrule
\end{tabular}
\end{table}

\end{document}